 \def\(#1){(\ref{#1})}

\documentstyle[prd,aps,eqsecnum]{revtex}

\begin{document}

\title{Noise Kernel in Stochastic Gravity and Stress Energy Bi-Tensor of 
       Quantum Fields in Curved Spacetimes}
\author{
Nicholas. G. Phillips
    \thanks{Electronic address: {\tt Nicholas.G.Phillips@gsfc.nasa.gov}} \\
{\small Raytheon ITSS, Laboratory for Astronomy and Solar Physics, Code 685,
             NASA/GSFC, Greenbelt, Maryland 20771}\\
B. L. Hu 
    \thanks{Electronic address: {\tt hub@physics.umd.edu}}\\
{\small Department of Physics, University of Maryland, College Park, Maryland 
20742-4111}}
\date{Submitted to Phys. Rev. D, Oct. 5, 2000}
\maketitle

\begin{abstract}
The noise kernel  is the vacuum expectation value of the  (operator-valued) 
stress-energy bi-tensor which describes the fluctuations of  a quantum field in
curved spacetimes.  It plays the role in stochastic semiclassical gravity based
on the Einstein-Langevin equation similar to the expectation value of the 
stress-energy tensor in semiclassical gravity based on the semiclassical
Einstein equation. According to the stochastic gravity program, this two point 
function (and by extension the higher order correlations in a hierarchy) of the 
stress energy
tensor possesses  precious statistical mechanical information of quantum fields
in curved spacetime and,  by the self-consistency required of Einstein's
equation, provides a probe into the coherence  properties of the gravity sector
(as measured by the higher order correlation functions of gravitons)
and the quantum and/or extended nature of spacetime.   It reflects the
medium energy (referring to Planck energy  as high energy) or mesoscopic
behavior of any viable theory of quantum gravity, including string theory. 
The stress energy bi-tensor could be the starting point for a new quantum field
theory constructed on spacetimes with extended structures. In the
coincidence limit we use the method of  point-separation to 
derive a  regularized  noise-kernel for a scalar field in general  curved 
spacetimes. 
It is useful for calculating quantum fluctuations of fields in modern
theories of structure formation and for  backreaction problems in the early
universe and black holes. 
One collorary of our finding is that for a massless conformal 
field the trace of the noise kernel identically vanishes. 
We outline how the general framework and results derived here 
can be used for the  calculation of noise kernels  for specific cases of 
physical interest such as the Robertson-Walker and Schwarzschild spacetimes.
\end{abstract}
\newpage



\section{Introduction}
\label{sec-ps-intro}

The central focus of this work is the noise kernel, which is the vacuum
expectation  value of the stress-energy bi-tensor for a quantum field in curved
spacetime \cite{BirDav,Full89,Wald94}.  It plays the role in stochastic 
semiclassical gravity 
\cite{stogra,MarVer} similar to the expectation value of the stress-energy 
tensor in semiclassical
gravity \cite{BirDav}. We believe that this two point function (and the 
hierarchy of
higher order correlation functions) of the stress energy tensor possesses 
precious
statistical mechanical information about  quantum fields  in curved spacetime 
and
reflects the low and medium energy (referring to Planck energy as high energy)
behavior of any viable theory of quantum gravity, including string theory. The
context for our investigation is stochastic gravity and the methodology is
point-separation. Let us first examine the context and then the methodology.

\subsection{Stochastic Gravity}

In semiclassical  gravity  the classical spacetime
(with metric $g_{ab}$) is driven by  the expectation value $\langle \rangle$  of the
stress energy  tensor $T_{ab}$  of a quantum field with respect to some quantum
state.  One  main task in the 70's was to obtain a regularized expression  for
this quantum source of  the semiclassical Einstein equation (SCE)  (e.g.,
\cite{CH87,CV94} and earlier work referred therein).  In stochastic 
semiclassical gravity of the 90's
\cite{stogra,MarVer} the additional effect of fluctuations of the stress energy
tensor \cite{KuoFor,PH97,PH00} enters which induces metric fluctuations in the
classical spacetime described by the Einstein-Langevin  equation  (ELE)
\cite{CH94,HM3}.      This stochastic term  measures the fluctuations of
quantum sources (e.g., arising from the difference of particles created in
neighboring histories \cite{CH94}) and  is intrinsically linked to the
dissipation in the dynamics of spacetime by a fluctuation -dissipation relation
\cite{HuSin,CV96}, which embodies the full backreaction effects of quantum
fields on classical spacetime.

The stochastic semiclassical Einstein equation, or Einstein-Langevin equation, 
takes on the form
\begin{eqnarray}
     G_{ab}[g] + \Lambda g_{ab}
        &=& 8\pi G ( {T_{ab}}^c +  {T_{ab}}^{qs})
           \nonumber \\
        T_{ab}^{qs}
        &\equiv& \langle T_{ab} \rangle_{q} +  T_{ab}^{s}
   \label{ELE}
\end{eqnarray}
where $G_{ab}$ is the Einstein tensor associated with $g_{ab}$ and $\Lambda, G$ 
are the
cosmological and Newton constants respectively. Here we use the superscripts c, 
s, q to 
denote classical, stochastic and quantum respectively.
The new term $T_{ab}^{s}= 2 \tau_{ab}$ which is of classical  
stochastic nature measures the  fluctuations of the energy momentum tensor 
of the quantum field. To see what $\tau_{ab}$ is, first define
\begin{equation}
{\hat t}_{ab}(x) \equiv {\hat T}_{ab}(x) - \langle {\hat T}_{ab}(x)\rangle \hat I
\label{that}
\end{equation}
which is a tensor operator measuring the deviations from the mean of the stress
energy tensor. We are interested in the correlation of these operators at
different spacetime points. Here we focus on the stress energy (operator-valued)
bi-tensor $\hat t_{ab} (x) \hat t_{c'd'}(y)$ defined at nearby points $(x, y)$.
metric.)
A bi-tensor is a geometric
object that has support at two separate spacetime points. In particular, it is
a rank two tensor  in the tangent space at $x$ (with unprimed indices) and in
the tangent space at $y$ (with primed indicies).

The noise kernel $N_{abc'd'}$ bitensor is defined as
\begin{equation}
4 N_{abc'd'} (x,y) \equiv {1\over 2} \langle \{ {\hat t}_{ab}(x), 
{\hat t}_{c'd'}(y) \} \rangle 
\label{D4}
\end{equation}
where $\{ \}$ means taking the symmetric product.   In the influence functional
(IF) \cite{if} or closed-time-path (CTP) effective action approach \cite{ctp} 
the
noise kernel appears in the real part of the influence action \cite{CH94,Banff}. 
\footnote{The derivation of noise kernel from IF or  CTP coarse-grained 
effective 
action is well-known from earlier work \cite{CH94,HuSin,CV96,MarVer}. Generally
speaking the effective  action  approach excels in the treatment of backreaction 
effects
of quantum fields on background spacetimes since it has self-consistency built 
in.
However, most calculations of the influence actions derived so far require some 
form of 
perturbative expansion, because it is difficult to obtain the influence 
functional in closed form,
(exceptions are only for  spacetimes with high symmetry,  and for special 
classes of coupling between the field 
and spacetime). We adopt the covariant point separation method because we can 
obtain a 
expression for the noise kernel in  general curved spacetime with explicit
dependence on the world function between two points and its covariant 
derivatives.  
This will be useful for later exploration of extended feature of spacetime.}
The noise kernel defines a real classical Gaussian stochastic symmetric tensor
field $\tau_{ab}$ which is characterized to lowest order by the following
relations,
\begin{equation}
\langle\tau_{ab}(x)\rangle_\tau=0,\ \ \ \ 
\langle \tau_{ab}(x) \tau_{c'd'}(y)\rangle_\tau= 
N_{abc'd'}(x,y),
\label{D6}
\end{equation}
where $\langle\,\rangle_\tau$ means taking a statistical average with respect
to the noise  distribution $\tau$ (for simplicity we don't consider higher
order correlations).  Since $\hat T_{ab}$ is self-adjoint, one can see that
$N_{abc'd'}$ is  symmetric, real, positive and semi-definite.  Furthermore, as a
consequence of (\ref{D4}) and the conservation law $\nabla^a \hat T_{ab}=0$,
this stochastic tensor $\tau_{ab}$
is divergenceless in the sense that $\nabla^a
\tau_{ab}=0$ is a deterministic zero field. Also 
 $g^{ab}{\tau}_{ab}(x) = 0$,  signifying that there is no  
stochastic correction to the trace anomaly (if $T_{ab}$ is traceless). 
(See \cite{MarVer}). Here  all covariant
derivatives are taken with respect to the background metric $g_{ab}$ which
is a solution of the semiclassical equations. Taking the statistical average of 
 (\ref{ELE}) with respect to the noise distribution $\tau$, 
as a consequence of the noise correlation relation (\ref{D6}),
\begin{equation}
          \langle T_{ab}^{qs}
          \rangle_\tau
        = \langle T_{ab} \rangle_q
\label{meant}
\end{equation}
we recover the semiclassical Einstein equation 
which is (\ref{ELE}) without the $T_{ab}^s$  term.
It is in this sense that we view semiclassical gravity as a mean field theory.

\subsection{Relation to Semiclassical and Quantum Gravity}

Stochastic semiclassical gravity thus ingrains a relation between
noise in quantum fields and metric fluctuations. While
the semiclassical regime describes the effect of a quantum  matter field
only through its mean value (e.g., vacuum expectation value), the 
stochastic regime includes the effect of fluctuations and correlations.
 We believe precious new information resides in the two-point functions of the
stress energy tensor which may reflect the finer structure of spacetime at a
scale when information provided by its mean value as source (semiclassical 
gravity) 
is no longer adequate.  
To appreciate this, it is perhaps instructive to examine the
distinction among these three  theories: stochastic gravity in relation to 
semiclassical and quantum  gravity \cite{MarVer,stogra}.  
The following observation will also bring
out two other related concepts of correlation (in the quantum field) and
coherence (in quantum gravity).

\subsubsection{Classical, Stochastic and Quantum}

For concreteness we consider the example of gravitational perturbations
$h_{ab}$ in a  background spacetime with metric $g_{ab}$ driven by the
expectation value of the energy momentum tensor of a scalar field $\Phi$, as
well as its fluctuations ${\hat t}_{ab}(x)$ \cite{HuSin,CV96}. Let us
compare the stochastic with the semiclassical and quantum equations  of motion
for the metric perturbation (weak but deterministic) field $h$. 
(This schematic representation was made by 
E. Verdaguer in \cite{MarVer}). The semiclassical equation is given by
\begin{equation}
\Box h =16\pi G \langle \hat T \rangle
\end{equation}
where $\langle \rangle$ denotes taking the quantum average (e.g., the vacuum
expectation value)  of the operator enclosed. Its solution can be written in
the form
\begin{equation}
h = \int C  \langle \hat T\rangle, ~~~~~
h_1h_2 = \int\int C _1C _2 \langle \hat T\rangle \langle\hat T\rangle.
\end{equation}
The quantum (Heisenberg) equation
\begin{equation}
\Box \hat h =16 \pi G  \hat T
\end{equation}
has solutions
\begin{equation}
\hat h = \int C  \hat T, ~~~~~
\langle \hat h_1 \hat h_2\rangle = \int \int C _1 C _2 \langle \hat T \hat 
	T \rangle_{\hat h, \hat \phi}
\end{equation}
where the average is taken  with respect to
the quantum fluctuations of both the gravitational ($ \hat g $) and 
the matter ($\hat \phi$) fields. Now for the stochastic equation, we have
\begin{equation}
\Box h =16 \pi G ( \langle \hat T \rangle + \tau)
\end{equation}
with solutions
\footnote{In this schematic form we have not displayed the homogeneous
 solution carrying the information of the (maybe random) initial condition. 
This solution will  exist in general, and may even be dominant if dissipation is 
weak .
When both the uncertainty in initial conditions and
the stochastic noise are taken into account, the Einstein - Langevin
formalism reproduces the exact graviton two point function, in the
linearized approximation. Of course, it fails to reproduce the expectation
value of observables which could not be written in terms of graviton
occupation numbers, and in this sense it falls short of full quantum gravity. 
This comment was made by  E. Calzetta to the author of \cite{stogra}.}

\begin{equation}
h = \int C  \langle \hat T \rangle + \int C  \tau, ~~~~~
h_1 h_2 = \int\int C _1 C _2 [ \langle \hat T \rangle \langle \hat T \rangle + 
(\langle \hat T \rangle \tau + \tau \langle \hat T \rangle) + \tau \tau]
\end{equation}
Now take the noise average $\langle \rangle_\tau$ . Recall that the noise
is defined in terms of the stochastic sources $\tau$ as
\begin{equation}
\langle \tau \rangle_\tau = 0, ~~~~ \langle \tau_1\tau_2\rangle_\tau \equiv 
	\langle \hat T_1 \hat T_2 \rangle - \langle \hat T_1\rangle
\langle \hat T_2\rangle
\end{equation}
we get
\begin{equation}
\langle h_1 h_2 \rangle_\tau = \int \int C _1 C _2 \langle \hat T \hat T \rangle_{\hat \phi}
\end{equation}
Note that the correlation of the energy momentum tensor appears just like in
the quantum case, but the average here is  over noise from quantum
fluctuations of the matter field alone. 
 
\subsubsection{Fluctuations, Correlations and Coherence}

Comparing the equations above depicting the semiclassical, stochastic and 
quantum regimes, \footnote{The observations in this section were first
made in \cite{stogra}} we see first that in the semiclassical case, the 
classical
metric correlations is given by the product of the vacuum expectation value of
the energy momentum tensor whereas in the quantum case it is given by 
the quantum average of the correlation of metric (operators) with
respect to the fluctuations in both the matter and the gravitational fields. In
the stochastic case the form is  closer to the quantum case except that the
quantum average is replaced by the noise average, and the average of the energy
momentum tensor is taken with respect only to the matter field. The important
improvement over semiclassical gravity is that it now carries information on
the correlation of the energy momentum tensor of the fields and its induced
metric fluctuations. Thus stochastic gravity contains information about the
correlation of fields (and the related phase information) which is absent in
semiclassical gravity. Here we have invoked the relation between {\it
fluctuations} and {\it correlations}, a variant form of the
fluctuation-dissipation relation. This feature moves stochastic gravity  closer
than semiclassical gravity to quantum gravity in that the correlation in
quantum field and geometry fully present in quantum gravity is partially
retained in stochastic gravity, and the background geometry has a way to sense
the correlation of the quantum fields through the noise term in the
Einstein-Langevin equation, which shows up as metric fluctuations.

By now we can see that `noise' as used in this more precise language and
broader context is not something one can arbitrarily assign  or relegate, as is
often done in ordinary discussions, but it has taken on a deeper meaning in
that it embodies the contributions of the higher correlation functions in the
quantum field. It holds the key to probing the quantum nature of spacetime
in this vein. We begin our studies here with the lowest order term, i.e., the 2 
point function
of the energy momentum tensor which contains the 4th order correlation of the
quantum field (or gravitons when they are considered as matter
source).\footnote{Although the Feynman- Vernon way can only accomodate Gaussian
noise of the matter fields and takes a simple form for linear coupling to the
background spacetime, the notion of noise can be made more general and precise.
For an example of more complex noise associated with more involved
backreactions arising from strong or nonlocal couplings, see Johnson and Hu
\cite{JohHu}} Progress is made now on how to characterize the higher order
correlation functions of an interacting field systematically from the
Schwinger-Dyson equations in terms of `correlation noise' \cite{cddn,StoBol},
after the BBGKY hierarchy. This may prove useful for a correlation dynamics
/stochastic semiclassical approach to quantum gravity \cite{stogra}.

Thus noise carries information about the correlations of the quantum field. One
can further link {\it correlation} in quantum fields to {\it coherence} in
quantum gravity. This stems from the self-consistency required in the
backreaction equations for the matter and spacetime sectors. The
Einstein-Langevin equation is only a partial (low energy) representation of the
complete theory of quantum gravity and fields. There, the coherence in the
geometry is related to the coherence in the matter field, as the complete
quantum description should be given by a coherent wave function of the combined
matter and gravity sectors.  Semiclassical gravity forsakes all the coherence
in the quantum gravity sector. Stochastic gravity captures only partial
coherence in  the quantum gravity sector via the correlations in the quantum
fields.
Since the degree of coherence can be measured in terms of correlations, our
strategy for the semiclassical stochastic gravity program  is to unravel the 
higher correlations of the matter field, starting
with the variance of the stress energy tensor and through its linkage with
gravity, retrieve whatever quantum attributes (partial coherence) of gravity 
left over
from the high energy behavior above the Planck scale.  Thus in this approach, 
focussing on
the noise kernel and the stress energy tensor two point function is our first
step beyond mean field (semiclassical gravity) theory  towards probing the full
theory of quantum gravity. 

\subsection{Point Separation and Noise Kernel}

So far we have explained the physical motivation for investigating the noise
kernel and  stress energy bi-tensor. We now turn to the methodology. 

In the light of the above discussions, the point separation scheme introduced
in the 60's by DeWitt  \cite{DeWitt65} will be well suited for our purpose
here.  It was brought to more popular use in the 70's  in the context of
quantum field theory in curved spacetimes \cite{DeWitt75,Chis76,Chis78}
as a means for obtaining a finite quantum stress tensor.  Since the stress
tensor is built from the product of a pair of field operators evaluated at a
single point, it is not well-defined. In this scheme, one introduces an
artificial separation of the single point $x$ to a pair of closely separated
points $x$ and $x'$. The problematic terms involving field products such as 
$\hat\phi(x)^2$ becomes $\hat\phi(x)\hat\phi(x')$, whose expectation value is
well defined. One then brings the two points back (taking the coincidence
limit) to identify the divergences present, which will then be removed
(regularization) or moved (by renormalizing the coupling constants), thereby
obtaining  a well-defined, finite stress tensor at a single point. In this
context point separation was used as a technique (many practitioners may still
view it as a trick, even a clumpsy one) for the purpose of identifying the
ultraviolet divergences.

By contrast, in our program, point separated expression of stress energy
bi-tensor have fundamental physical meaning as it  contains information  on
the fluctuations and correlation of quantum fields,  and by consistency with
the gravity sector, can provide a probe into the coherent properties of quantum
spacetimes. Taking this view, we may also gain a new perspective on ordinary
quantum field theory defined on single points:
The coincidence limit depicts the low energy limit of the full quantum theory
of matter and spacetimes. Ordinary (pointwise) quantum field theory, classical
general relativity and  semiclassical gravity are the lowest levels of
approximations and should be viewed not as fundamental,  but only as effective
theories.  As such, even the  way how the conventional point-defined field 
theory 
emerges from the full theory when the two points (e.g., $x$ and $y$ in the noise 
kernel) are
brought together is interesting. For example, one can ask if there will also be
a quantum to classical transition in spacetime accompanying the  coincident
limit? Certain aspects like decoherence has been investigated before (see,
e.g., \cite{PazSin}), but here the non-local  structure of spacetime and their
impact on quantum field theory become the central issue. (This may also be a
relevant issue in noncommutative geometry). The point-wise limit of field theory
of course has ultraviolet divergences and requires regularization. A new 
viewpoint
towards regularization evolved from this perspective of treating  conventional 
pointwise field theory as an effective theory in the coincident limit of the 
point-separated 
theory of extended spacetime. This is discussed in our 
recent paper on fluctuations of the vacuum energy density and  the validity of 
semiclassical gravity \cite{PH00}.

The paper is organized as follows: In Section \ref{sec-ps-psreview}, we review
the method of point separation. In Section III, we discuss the procedures 
for dealing with the quantum stress tensor bi-operator at two
separated points and the noise kernel. 
We derive a general expression for the noise kernel in terms of the quantum 
field's Green
function and its covariant derivatives up to the fourth order . (The stress
tensor involves up to two covariant derivatives.) This result holds for $x\ne
y$ without recourse to renormalization of the Green function, showing that
$N_{abc'd'}(x,y)$ is always finite for $x\ne y$ (and off the light cone for
massless theories). Using this result, we show there is no stochastic
correction to the trace anomaly, in agreement with results arrived at in
\cite{CH94,CV96}.  In Section \ref{sec-ps-reg} we briefly review how 
``modified''
point  separation \cite{Wald75,ALN77,Wald78} is used to get a regularized Green
function.  This amounts to subtracting a locally determined ``Hadamard ansatz''
from the Green function. With this we show how to compute the fluctuations 
of the stress tensor, culminating in a general expression of the noise kernel 
and its
coincident form for an arbitrary curved spacetime. Paper II in this series  will
apply these formulas to ultrastatic metrics including the Einstein universe 
\cite{PH97},
hot flat space and optical Schwarzschild spacetimes. Paper III treats noise 
kernels
in the Robertson-Walker universe and Schwarzschild black holes, from which 
structure formation
from quantum fluctuations \cite{CH95} and backreaction of Hawking radiation on 
the black hole
spacetime \cite{HRS} can be studied. Finally, we intend in Paper IV to use the 
symbolic routine
to derive or check on analytic expressions for the stress-energy bi-tensor
in de Sitter and anti-de Sitter spacetimes. (Martin, Roura and Verdaguer 
\cite{MRV}
have obtained analytic expressions for Minkowski and restricted cases of de 
Sitter spaces.)
The former is necessary for scrutinizing 
primordial fluctuations in the cosmic background radiation while
the latter is related to black hole phase transition and  AdS/CFT issues in 
string theory.


\section{Point Separation}
\label{sec-ps-psreview}

We start with a short  overview of point separation in this section. 
This paper will present the general schema and the results for the coincident 
limit of
the noise kernel. The details, along with how they are carried out using 
symbolic 
computation, will be described in a separate paper by one of us \cite{NPsc}. 
Here we first review the construction of the stress tensor and 
then derive the symmetric stress tensor two point function, the
noise kernel, in terms of the Wightman Green function. Our
result is completely general for the case of a free scalar field
(the separation is maintained throughout)  and thus  can be used
with or without regularizing the Green function. Of course, without
use of a regularized Green function, the $y \rightarrow x$ limit is divergent.
We also give the noise kernel for a massless conformally coupled free scalar
field on a four dimensional manifold. We then compute the trace at both
points and find this double trace vanishes identitically for the massless
conformal case; implying that there is no noise associated with the trace 
anomaly.
This result holds separate from the issue of regularization.

Having the point-separated expression for the noise kernel and using the
regularized Green function, as is done for the stress tensor,
we obtain an expression for which the $y\rightarrow x$ limit is meaningful.
We use the so-call ``modified'' point separation prescription
\cite{Wald75,ALN77,Wald78}. In this procedure,
the naive Green function is rendered finite by assuming the divergences
present for $y \rightarrow x$ are state independent and can be removed by
subtraction of a Hadamard form. We review this prescription and use it 
to obtain a definition of the noise kernel for which the $y \rightarrow x$ limit 
is meaningful. \footnote{We know from work in the 70's in quantum field 
in curved spacetime \cite{BirDav,Full89,Wald94} that there are several 
regularization 
methods developed for the removal of ultraviolet divergences in the stress 
energy
tensor. Their mutual relations are known, and discrepancies explained. 
This formal structure of regularization schemes for  quantum fields in curved 
spacetime should remain intact as we apply them to the regularization
of the noise kernel in general curved spacetimes. Specific considerations will 
of course
enter for each method. But for the methods we have employed so far, such as  
zeta-function,
point separation,  smeared-field \cite{PH97,PH00} applied to simple cases 
(Casimir, Einstein, thermal fields) there is no new  inconsistency or 
discrepancy.}

\subsection{n-tensors and end-point expansions}

An object like the Green function $G(x,y)$ is an example of a
{\em bi-scalar}: it transforms as scalar at both points $x$ and $y$.
We can also define a {\em bi-tensor}\, 
$T_{a_1\cdots a_n\,b'_1\cdots b'_m}(x,y)$:
upon a coordinate transformation, this transforms as a rank $n$ tensor at 
$x$ and a rank $m$ tensor at $y$. We will extend this up to a 
{\em quad-tensor}\, 
$T_{a_1\cdots a_{n_1}\,b'_1\cdots b'_{n_2}\,
    c''_1\cdots c''_{n_3}\,d'''_1\cdots d'''_{n_4}}$ 
which has support at
four points $x,y,x',y'$, transforming as rank $n_1,n_2,n_3,n_4$ tensors
at each of the four points. This also sets the notation we will use:
unprimed indices referring to the tangent space constructed
above $x$, single primed indices to
$y$, double primed to $x'$ and triple primed to $y'$.
For each point, there is the covariant derivative $\nabla_a$ at that point.
Covariant derivatives at different points commute and the covariant
derivative at, say, point $x'$, does not act on a bi-tensor defined at,
say,  $x$ and $y$:
\begin{equation}
T_{ab';c;d'} = T_{ab';d';c} \quad {\rm and } \quad
T_{ab';c''} = 0.
\end{equation}
To simplify notation, henceforth we will eliminate the semicolons
after the first one for multiple covariant derivatives at multiple points.

Having objects defined at different points, the {\rm coincident limit} is
defined as evaluation ``on the diagonal'',
in the sense of the spacetime support of the function or tensor,
and the usual shorthand
$\left[ G(x,y) \right] \equiv G(x,x)$ is used. This extends to $n$-tensors
as
\begin{equation}
\left[ T_{a_1\cdots a_{n_1}\,b'_1\cdots b'_{n_2}\,
    c''_1\cdots c''_{n_3}\,d'''_1\cdots d'''_{n_4}} \right] = 
T_{a_1\cdots a_{n_1}\,b_1\cdots b_{n_2}\,
    c_1\cdots c_{n_3}\,d_1\cdots d_{n_4}},
\end{equation}
{\it i.e.}, this becomes a rank $(n_1+n_2+n_3+n_4)$ tensor at $x$.
The multi-variable chain rule relates covariant derivatives acting at
different points, when we are interested in the coincident limit:
\begin{equation}
\left[ T_{a_1\cdots a_m \,b'_1\cdots b'_n} \right]\!{}_{;c} = 
\left[ T_{a_1\cdots a_m \,b'_1\cdots b'_n;c} \right] +
\left[ T_{a_1\cdots a_m \,b'_1\cdots b'_n;c'} \right].
\label{ref-Synge's}
\end{equation}
This result is referred to as {\em Synge's theorem} in this context.
(We  follow Fulling's \cite{Full89} discussion.)

The bi-tensor of {\em parallel transport}\, $g_a{}^{b'}$ is defined such
that when it acts on a vector $v_{b'}$ at $y$, it parallel transports
the vector along the geodesics connecting $x$ and $y$. This allows us to
add vectors and tensors defined at different points. We cannot directly add a
vector $v_a$ at $x$ and vector $w_{a'}$ at $y$. But by using $g_a{}^{b'}$,
we can construct the sum
$v^a + g_a{}^{b'} w_{b'}$.
We will also need the obvious property $\left[ g_a{}^{b'} \right] = g_a{}^b$.

The main bi-scalar we need for this work is the {\em world function}
$\sigma(x,y)$. This is defined as a half of the square of the geodesic
distance between the points $x$ and $y$.
It satisfies the equation
\begin{equation}
\sigma = \frac{1}{2} \sigma^{;p} \sigma_{;p}
\label{define-sigma}
\end{equation}
Often in the literature, a covariant derivative is implied when the
world function appears with indices: $\sigma^a \equiv \sigma^{;a}$,
{\it i.e.}taking the covariant derivative at $x$, while $\sigma^{a'}$ means
the covariant derivative at $y$.
This is done since the vector $-\sigma^a$ is the tangent vector to the
geodesic with length equal the distance between $x$ and $y$. 
As $\sigma^a$ records information about distance and direction for the two
points  this makes it ideal for constructing a series expansion of
a bi-scalar.  The {\em end point} expansion of a bi-scalar $S(x,y)$ is
of the form
\begin{equation}
S(x,y) = A^{(0)} + \sigma^p A^{(1)}_p
+ \sigma^p \sigma^q A^{(2)}_{pq}
+ \sigma^p \sigma^q  \sigma^r A^{(3)}_{pqr}
+ \sigma^p \sigma^q  \sigma^r \sigma^s A^{(4)}_{pqrs} + \cdots
\label{general-endpt-series}
\end{equation}
where, following our convention, the expansion tensors 
$A^{(n)}_{a_1\cdots a_n}$ with  unprimed indices have support at $x$
and hence the name end point expansion. Only the
symmetric part of these tensors  contribute to the expansion.
For the purposes of multiplying series expansions it is convenient to
separate the distance dependence from the direction dependence. This
is done by introducing the unit vector $p^a = \sigma^a/\sqrt{2\sigma}$.
Then the series expansion can be written
\begin{equation}
S(x,y) = A^{(0)} + \sigma^{\frac{1}{2}} A^{(1)}
+ \sigma  A^{(2)}
+ \sigma^{\frac{3}{2}} A^{(3)}
+ \sigma^2 A^{(4)} + \cdots
\end{equation}
The expansion scalars are related to the expansion tensors via\\
$A^{(n)} = 2^{n/2} A^{(n)}_{p_1\cdots p_n}
p^{p_1}\cdots p^{p_n}$.

The last object we need  is the {\em VanVleck-Morette}
determinant $D(x,y)$, defined as
$D(x,y) \equiv -\det\left( -\sigma_{;ab'} \right)$.
The related bi-scalar
\begin{equation}
{\Delta\!^{1/2}} = \left( \frac{D(x,y)}{\sqrt{g(x) g(y)}}\right)^\frac{1}{2}
\end{equation}
satisfies the equation
\begin{equation}
{\Delta\!^{1/2}}\left(4-\sigma_{;p}{}^p\right) - 2{\Delta\!^{1/2}}_{\,\,;p}\sigma^{;p} = 0
\label{define-VanD}
\end{equation}
with the boundary condition $\left[{\Delta\!^{1/2}}\right] = 1$.

Further details on these objects and discussions of the definitions and
properties are contained in \cite{Chis76} and \cite{NPsc}. 
There it is shown how the defining equations for $\sigma$ and ${\Delta\!^{1/2}}$
are used to determine the coincident limit expression for the various covariant
derivatives of the world function
($\left[ \sigma_{;a}\right]$, $\left[ \sigma_{;ab}\right]$, {\it etc.})
and how the defining differential equation for ${\Delta\!^{1/2}}$ can be used
to determine the series expansion of ${\Delta\!^{1/2}}$.
We show how the expansion tensors $A^{(n)}_{a_1\cdots a_n}$
are determined in terms of the coincident limits of covariant
derivatives of the bi-scalar $S(x,y)$.  Ref.  \cite{NPsc} details how
point separation can be implemented on the computer to provide
easy access to a wider range of applications involving higher
derivatives of the curvature tensors. We will say a few words about this to end
this section.

\subsection{Symbolic Computation}

Since the noise kernel involves up to four covariant derivatives, we
expand the Green function out to fourth order in the distance between
the points in an end-point expansion. One of the main advantages of this
approach is that it readily lends itself to implementation in a symbolic
computing environment \cite{Wolfram,MathTensor}. 
So the task becomes one of developing the
necessary series expansions of the various geometric objects, including
the conformal transformation properties. From the symbolic computational
viewpoint, the key to this procedure is to recognize that all the objects
for which a series expansion are needed are defined by first or second
order covariant differential equations. This allows us to develop
recursive algorithms for the symbolic manipulations. The lowest order
terms in the series expansions are known  \cite{Chis76,Page82}. 
For example,  Christensen's work on the stress tensor
goes to second order in $\sigma^a$ (ending up creating results spanning a page 
and a half in his manuscript). The noise kernel results get much longer.
(Some expansion coefficients end up over 400 terms in length.)  
\footnote{The final expressions reaching  such a length defy direct inspection
check.  However, since the algorithms are recursive,  agreement  at the lower 
orders
extends to higher orders for the correctness of the complete result.  For 
insurance, 
some of the larger expressions are derived by two independent methods and 
verified 
to be the same.}

We have developed a systematic set of computer code, using MathTensor
\cite{MathTensor},  running in the Mathematica \cite{Wolfram} environment on a
workstation,  to carry out much of the work contained in here and the following
papers in this series. We use it to compute the fluctuations of the quantum
stress tensor for scalar fields from a given form of the metric and an analytic
expression for the Green function. We can also deal with cases where the Green
function is only known in an analytic form in a conformally related spacetime.
No recourse is made to numerical methods; the results are exact up to the
accuracy of the given analytic form of the Green  function 
{\em before} regularization. The code itself carries out the
regularization via the ``modified'' point separation prescription.

\section{Stress Energy Bi-Tensor Operator and Noise Kernel}
\label{sec-ps-stress-tensor}

Even though we believe that the point-separated results are more basic in the
sense that it reflects a deeper structure of the quantum theory of spacetime,
we will nevertheless start with  quantities defined at one point because they
are more familiar in conventional quantum field theory. We will use point
separation to introduce the biquantities. The key issue here is thus the
distinction between point-defined ({\it pt}) and point-separated ({\it bi}) 
quantities.

For  a free classical scalar field with the action
\begin{equation}
S[\phi] = -\frac{1}{2}\int\left(
{m^2}\,{{\phi }^2} + {{\phi }^2}\,R\,\xi  + {\phi {}_;{}_{p}}
\,{\phi {}^;{}^{p}}
\right)\sqrt{g}\,d^4x.
\end{equation}
the classical stress energy tensor  is
\begin{eqnarray}
T_{ab}(x) &\equiv& \frac{2}{\sqrt{g(x)}}
            \frac{\delta S[\phi]}{\delta g^{ab}(x)}\cr\cr
&=&
\left( 1 - 2\,\xi  \right) \,{\phi {}_;{}_{a}}\,{\phi {}_;{}_{b}}
 + \left(2\,\xi -{1\over 2} \right) \,{\phi {}_;{}_{p}}
\,{\phi {}^;{}^{p}}\,{g{}_{a}{}_{b}}
+ 2\xi\,\phi \, \,\left({\phi {}_;{}_{p}{}^{p} -{\phi {}_;{}_{a}{}_{b}}}
\,{g{}_{a}{}_{b}} \right)  \cr
&&+ {{\phi }^2}\,\xi \,
\left({R{}_{a}{}_{b}} - {1\over 2}{ R\,{g{}_{a}{}_{b}}  }
 \right) 
 - \frac{1}{2}{{m^2}\,{{\phi }^2}\,{g{}_{a}{}_{b}}}
\label{ref-define-classical-emt}
\end{eqnarray}
When we make the transition to quantum field theory,
we promote the field $\phi(x)$ to a field operator $\hat\phi(x)$.
The fundamental problem of defining a quantum operator for the 
stress tensor is immediately visible: the field operator 
appears quadratically. Since $\hat\phi(x)$ is an operator-valued
distribution, products at a single point are not well-defined.
But if the product is point separated ($\hat\phi^2(x) \rightarrow
\hat\phi(x)\hat\phi(x')$), they are finite and well-defined.

Let us first seek a point-separated extension of these classical quantities and
then consider the quantum field operators. Point separation is symmetrically
extended to products of covariant derivatives of the field according to
\begin{eqnarray}
\left({\phi {}_;{}_{a}}\right)\left({\phi {}_;{}_{b}}\right) &\rightarrow&
\frac{1}{2}\left(
g_a{}^{p'}\nabla_{p'}\nabla_{b}+g_b{}^{p'}\nabla_a\nabla_{p'}
\right)\phi(x)\phi(x'), \\
\phi \,\left({\phi {}_;{}_{a}{}_{b}}\right) &\rightarrow&
\frac{1}{2}\left(
\nabla_a\nabla_b+g_a{}^{p'}g_b{}^{q'}\nabla_{p'}\nabla_{q'}
\right)\phi(x)\phi(x').
\end{eqnarray}
The bi-vector of parallel displacement
$g_a{}^{a'}(x,x')$ is included so that we may 
have objects that are rank 2 tensors
at $x$ and scalars at $x'$.

To carry out  point separation on (\ref{ref-define-classical-emt}), we first
define the differential operator
\begin{eqnarray}
{\cal T}_{ab} &=&
  \frac{1}{2}\left(1-2\xi\right)
   \left(g_a{}^{a'}\nabla_{a'}\nabla_{b}+g_b{}^{b'}\nabla_a\nabla_{b'}\right)
+ \left(2\xi-\frac{1}{2}\right)
     g_{ab}g^{cd'}\nabla_c\nabla_{d'} \cr
&&
- \xi
     \left(\nabla_a\nabla_b+g_a{}^{a'}g_b{}^{b'}\nabla_{a'}\nabla_{b'}\right)
+ \xi g_{ab}
     \left(\nabla_c\nabla^c+\nabla_{c'}\nabla^{c'}\right) \cr
&&
+\xi\left(R_{ab} - \frac{1}{2}g_{ab}R\right)
    -\frac{1}{2}m^2 g_{ab}
\label{PSNoise-emt-diffop}
\end{eqnarray}
from which we obtain the classical stress tensor as
\begin{equation}
T_{ab}(x) = \lim_{x' \rightarrow x} {\cal T}_{ab}\phi(x)\phi(x').
\end{equation}
That the classical tensor field no longer appears as a product of scalar 
fields at a single point allows a smooth transition
to the quantum tensor field. From the viewpoint of the stress tensor, 
the separation of points is an artificial construct so when promoting the 
classical field to a
quantum one, neither point should be favored. The product of field
configurations is taken to be the symmetrized operator product,
denoted by curly brackets:
\begin{equation}
\phi(x)\phi(y) \rightarrow \frac{1}{2}
 \left\{{\hat\phi(x)},{\hat\phi(y)}\right\}
= \frac{1}{2}\left( {\hat\phi(x)} {\hat\phi(y)} +
                    {\hat\phi(y)} {\hat\phi(x)}
\right)
\end{equation}
With this, the point separated stress energy tensor operator is defined as
\begin{equation}
\hat T_{ab}(x,x') \equiv \frac{1}{2}
{\cal T}_{ab}\left\{\hat\phi(x),\hat\phi(x')\right\}.
\label{PSNoise-emt-define}
\end{equation}
While the classical stress tensor was defined at the coincidence limit
$x'\rightarrow x$, we cannot attach any physical meaning to the quantum stress 
tensor at one point
until the issue of regularization is dealt with, which will happen in the next 
section.
For now,  we will maintain point separation so as to have a mathematically 
meaningful operator. 

The expectation value of the point-separated stress tensor can now be
taken. This amounts to replacing the field operators by their expectation
value, which is given by the Hadamard (or Schwinger) function
\begin{equation}
{G^{(1)}}(x,x') =
     \langle\left\{{\hat\phi(x)},{\hat\phi(x')}\right\}\rangle.
\end{equation}
and the point-separated stress tensor is defined as
\begin{equation}
\langle \hat T_{ab}(x,x') \rangle = 
\frac{1}{2} {\cal T}_{ab}{G^{(1)}}(x,x')
\label{ref-emt-PSdefine}
\end{equation}
where, since ${\cal T}_{ab}$ is a differential operator, it can be taken
``outside'' the expectation value. The expectation value of the point-separated
quantum stress tensor for a free, massless ($m=0$)  conformally
coupled ($\xi=1/6$) scalar field on a four dimension spacetime with 
scalar curvature $R$ is
\begin{eqnarray}
\langle \hat T_{ab}(x,x') \rangle &=&
  \frac{1}{6}\left( {g{}^{p'}{}_{b}}\,{{G^{(1)}}{}_;{}_{p'}{}_{a}}
 + {g{}^{p'}{}_{a}}\,{{G^{(1)}}{}_;{}_{p'}{}_{b}} \right)
 -\frac{1}{12} {g{}^{p'}{}_{q}}\,{{G^{(1)}}{}_;{}_{p'}{}^{q}}
\,{g{}_{a}{}_{b}} \cr 
&& -\frac{1}{12}\left( {g{}^{p'}{}_{a}}\,{g{}^{q'}{}_{b}}
\,{{G^{(1)}}{}_;{}_{p'}{}_{q'}} + {{G^{(1)}}{}_;{}_{a}{}_{b}} \right)
 +\frac{1}{12}\left( \left( {{G^{(1)}}{}_;{}_{p'}{}^{p'}}
 + {{G^{(1)}}{}_;{}_{p}{}^{p}} \right) \,{g{}_{a}{}_{b}} \right) \cr 
&& +\frac{1}{12} {G^{(1)}}\,
\left({R{}_{a}{}_{b}} -{1\over 2} R\,{g{}_{a}{}_{b}} \right)
\end{eqnarray}


\subsection{Finiteness of Noise Kernel}

We now turn our attention to the noise kernel. First introduced in the context
of stochastic semiclassical gravity it is the symmetrized product of the (mean
subtracted) stress tensor operator:
\begin{eqnarray}
8 N_{ab,c'd'}(x,y) &=&
\langle \left\{ 
		\hat T_{ab}(x)-\langle \hat T_{ab}(x)\rangle,
	\hat T_{c'd'}(y)-\langle \hat T_{c'd'}(y) \rangle
\right\} \rangle \cr
&=&
\langle \left\{ \hat T_{ab}(x),\hat T_{c'd'}(y) \right\} \rangle
-2 \langle \hat T_{ab}(x)\rangle\langle \hat T_{c'd'}(y) \rangle
\end{eqnarray}
Since $\hat T_{ab}(x)$ defined at one point can be ill-behaved as it is 
generally divergent,
one can question the soundness of these quantities. But as will be shown later, 
the noise kernel
is finite for $y\neq x$. All field operator
products present in the first expectation value that could be
divergent are canceled by similar products in the second term.
We will replace each of the stress tensor operators in the above expression for 
the noise kernel
by their point separated versions,
effectively separating the two points $(x,y)$ into the four points
$(x,x',y,y')$. This will allow us to express the noise kernel in terms
of a pair of differential operators acting on a combination of
four and two point functions. Wick's theorem will allow the four
point functions to be re-expressed in terms of two point functions. From this
we see that all possible divergences for $y\neq x$ will cancel.
When the coincident limit is taken divergences do occur. The above procedure
will allow us to isolate the divergences and obtain a finite result.

Taking the point-separated quantities as more basic, one should replace each of 
the stress tensor operators in the above with the corresponding point separated 
version (\ref{PSNoise-emt-define}),
with ${\cal T}_{ab}$ acting at $x$ and $x'$ and ${\cal T}_{c'd'}$ acting
at $y$ and $y'$. In this framework the noise kernel is defined as
\begin{equation}
8 N_{ab,c'd'}(x,y) =
   \lim_{x'\rightarrow x}\lim_{y'\rightarrow y}
   {\cal T}_{ab} {\cal T}_{c'd'}\, G(x,x',y,y')
\end{equation}
where the four point function is
\begin{eqnarray}
G(x,x',y,y') &=& \frac{1}{4}\left[
\langle\left\{\left\{{\hat\phi(x)},{\hat\phi(x')}\right\},\left\{{\hat\phi(y)}
,{\hat\phi(y')}\right\}\right\}\rangle
\right. \cr\cr&&\hspace{1cm}\left.
  -2\,\langle\left\{{\hat\phi(x)},{\hat\phi(x')}\right\}\rangle
  \langle\left\{{\hat\phi(y)},{\hat\phi(y')}\right\}\rangle \right].
\label{PSNoise-G4a}
\end{eqnarray}
We assume the pairs $(x,x')$ and $(y,y')$ are each 
within their respective Riemann normal coordinate
neighborhoods so as to avoid problems
that possible geodesic caustics might be present. When we later
turn our attention to computing the limit $y\rightarrow x$, after
issues of regularization are addressed, we will want to assume all four
points are within the same Riemann normal coordinate neighborhood.

Wick's theorem, for the case of free fields which we
are considering, gives the simple product four point function in terms
of a sum of products of Wightman functions (we use the shorthand notation
$G_{xy}\equiv G_{+}(x,y) = \langle{\hat\phi(x)}\,{\hat\phi(y)}\rangle$):
\begin{equation}
\langle{\hat\phi(x)}\,{\hat\phi(y)}\,{\hat\phi(x')}\,{\hat\phi(y')}\rangle =
{G_{xy'}}\,{G_{yx'}} + {G_{xx'}}\,{G_{yy'}} + {G_{xy}}\,{G_{x'y'}}
\end{equation}
Expanding out the anti-commutators in (\ref{PSNoise-G4a}) and applying
Wick's theorem, the four point function becomes
\begin{equation}
G(x,x',y,y')  =
{G_{xy'}}\,{G_{x'y}} + {G_{xy}}\,{G_{x'y'}} + {G_{yx'}}\,{G_{y'x}} + {G_{yx}}
\,{G_{y'x'}}
\end{equation}
We can now easily see that the noise kernel defined via this
function is indeed well defined for the limit $(x',y')\rightarrow (x,y)$:
\begin{equation}
G(x,x,y,y) = 2\,\left( {{{G^2_{xy}}}} + {{{G^2_{yx}}}} \right) ,
\end{equation}
>From this we can see that the noise kernel is also well defined for $y \neq x$;
any divergence present in the first expectation value of
(\ref{PSNoise-G4a}) have been cancelled by those present in the
pair of Green functions in the second term.

\subsection{Explicit Form of the Noise Kernel}

We will let the points  separated for a while so we can keep tract of which
covariant derivative acts on which arguments of which Wightman function. 
As an example
(the complete calculation is quite long), consider the
result of the first set of covariant derivative operators in the
differential operator (\ref{PSNoise-emt-diffop}), from both
${\cal T}_{ab}$ and ${\cal T}_{c'd'}$, acting on $G(x,x',y,y')$:
\begin{eqnarray}
&&\frac{1}{4}\left(1-2\xi\right)^2
   \left(g_a{}^{p''}\nabla_{p''}\nabla_{b}+
         g_b{}^{p''}\nabla_{p''}\nabla_{a}\right)\cr
&&\hspace{17mm}\times
   \left(g_{c'}{}^{q'''}\nabla_{q'''}\nabla_{d'}
        +g_{d'}{}^{q'''}\nabla_{q'''}\nabla_{c'}\right)
    G(x,x',y,y')
\end{eqnarray}
(Our notation is that $\nabla_a$ acts at $x$, $\nabla_{c'}$ at $y$,
$\nabla_{b''}$ at $x'$ and $\nabla_{d'''}$ at $y'$).
Expanding out the differential operator above, we can determine which
derivatives act on which Wightman function:
\begin{eqnarray}
{{{{\left( 1 - 2\,\xi  \right) }^2}}\over 4} &\times & \left[
    {g{}_{c'}{}^{p'''}}\,{g{}^{q''}{}_{a}}
 \left( {{G_{xy'}}{}_;{}_{b}{}_{p'''}}\,{{G_{x'y}}{}_;{}_{q''}{}_{d'}}
 + {{G_{xy}}{}_;{}_{b}{}_{d'}}\,{{G_{x'y'}}{}_;{}_{q''}{}_{p'''}}
 \right.\right. \cr
&&  \hspace{20mm} + \left. {{G_{yx'}}{}_;{}_{q''}{}_{d'}}
\,{{G_{y'x}}{}_;{}_{b}{}_{p'''}} + {{G_{yx}}{}_;{}_{b}{}_{d'}}
\,{{G_{y'x'}}{}_;{}_{q''}{}_{p'''}} \right) \cr
&&+ {g{}_{d'}{}^{p'''}}\,{g{}^{q''}{}_{a}}
 \left( {{G_{xy'}}{}_;{}_{b}{}_{p'''}}\,{{G_{x'y}}{}_;{}_{q''}{}_{c'}}
 + {{G_{xy}}{}_;{}_{b}{}_{c'}}\,{{G_{x'y'}}{}_;{}_{q''}{}_{p'''}} \right. \cr
&&  \hspace{20mm} + \left. {{G_{yx'}}{}_;{}_{q''}{}_{c'}}
\,{{G_{y'x}}{}_;{}_{b}{}_{p'''}} + {{G_{yx}}{}_;{}_{b}{}_{c'}}
\,{{G_{y'x'}}{}_;{}_{q''}{}_{p'''}} \right) \cr
&&+ {g{}_{c'}{}^{p'''}}\,{g{}^{q''}{}_{b}}
 \left( {{G_{xy'}}{}_;{}_{a}{}_{p'''}}\,{{G_{x'y}}{}_;{}_{q''}{}_{d'}}
 + {{G_{xy}}{}_;{}_{a}{}_{d'}}\,{{G_{x'y'}}{}_;{}_{q''}{}_{p'''}} \right. \cr
&&  \hspace{20mm} + \left. {{G_{yx'}}{}_;{}_{q''}{}_{d'}}
\,{{G_{y'x}}{}_;{}_{a}{}_{p'''}} + {{G_{yx}}{}_;{}_{a}{}_{d'}}
\,{{G_{y'x'}}{}_;{}_{q''}{}_{p'''}} \right) \cr
&&+ {g{}_{d'}{}^{p'''}}\,{g{}^{q''}{}_{b}}
 \left( {{G_{xy'}}{}_;{}_{a}{}_{p'''}}\,{{G_{x'y}}{}_;{}_{q''}{}_{c'}}
 + {{G_{xy}}{}_;{}_{a}{}_{c'}}\,{{G_{x'y'}}{}_;{}_{q''}{}_{p'''}} \right. \cr
&&  \hspace{20mm} + \left.\left. {{G_{yx'}}{}_;{}_{q''}{}_{c'}}
\,{{G_{y'x}}{}_;{}_{a}{}_{p'''}} + {{G_{yx}}{}_;{}_{a}{}_{c'}}
\,{{G_{y'x'}}{}_;{}_{q''}{}_{p'''}} \right) \right]
\end{eqnarray}
If we now  let $x'\rightarrow x$ and $y' \rightarrow y$
the contribution to the noise kernel is (including the factor of
$\frac{1}{8}$ present in the definition of the noise kernel):
\begin{eqnarray}
&&\frac{1}{8}\left\{ {{\left( 1 - 2\,\xi  \right) }^2}
\,\left( {{G_{xy}}{}_;{}_{a}{}_{d'}}\,{{G_{xy}}{}_;{}_{b}{}_{c'}}
 + {{G_{xy}}{}_;{}_{a}{}_{c'}}\,{{G_{xy}}{}_;{}_{b}{}_{d'}}
 \right)  \right. \cr
&&\hspace{20mm} \left. + {{\left( 1 - 2\,\xi  \right) }^2}
\,\left( {{G_{yx}}{}_;{}_{a}{}_{d'}}\,{{G_{yx}}{}_;{}_{b}{}_{c'}}
 + {{G_{yx}}{}_;{}_{a}{}_{c'}}\,{{G_{yx}}{}_;{}_{b}{}_{d'}} \right)  \right\}
\end{eqnarray}
That this term can be written as the sum of a part involving $G_{xy}$ and
one involving $G_{yx}$ is a general property of the entire noise kernel. It thus
takes the form
\begin{equation}
N_{abc'd'}(x,y) = N_{abc'd'}\left[ G_{+}(x,y)\right]
                + N_{abc'd'}\left[ G_{+}(y,x)\right].
\end{equation}
We will present the form of the functional $N_{abc'd'}\left[ G \right]$
shortly. First we note, for $x$ and $y$ time-like separated, the above
split of the noise kernel allows us to express it
in terms of the Feynman (time ordered) Green function $G_F(x,y)$ and
the Dyson (anti-time ordered) Green function $G_D(x,y)$:
\begin{equation}
N_{abc'd'}(x,y) = N_{abc'd'}\left[ G_F(x,y)\right]
                + N_{abc'd'}\left[ G_D(x,y)\right]
\label{noiker}
\end{equation}
\footnote{This can be connected  with the zeta function approach to this 
problem \cite{PH97} as follows: Recall when the quantum stress tensor
fluctuations determined in the Euclidean section is analytically continued back
to Lorentzian signature ($\tau \rightarrow i t$), the time ordered product
results. On the other hand, if the continuation is $\tau \rightarrow -i t$, the
anti-time ordered product results. With this in mind, the noise kernel is seen
to be related to the quantum stress tensor fluctuations derived via the
effective action as
\begin{equation}
16 N_{abc'd'} =
   \left.\Delta T^2_{abc'd'}\right|_{t=-i\tau,t'=-i\tau'}
 + \left.\Delta T^2_{abc'd'}\right|_{t= i\tau,t'= i\tau'}.
\end{equation}
 }
The complete form of the functional $N_{abc'd'}\left[ G \right]$ is
%
%
%
\begin{mathletters}
\label{ref-noise-kernel}
\begin{equation}
 N_{abc'd'}\left[ G \right]  = 
    \tilde N_{abc'd'}\left[ G \right]
  + g_{ab}   \tilde N_{c'd'}\left[ G \right]
 + g_{c'd'} \tilde N'_{ab}\left[ G \right]
 + g_{ab}g_{c'd'} \tilde N\left[ G \right]
\label{general-noise-kernel}
\end{equation}
with
\begin{eqnarray}
8 \tilde N_{abc'd'} \left[ G \right]&=& 
%
{{\left(1-2\,\xi\right)}^2}\,\left( G{}\!\,_{;}{}_{c'}{}_{b}\,
     G{}\!\,_{;}{}_{d'}{}_{a} + 
    G{}\!\,_{;}{}_{c'}{}_{a}\,G{}\!\,_{;}{}_{d'}{}_{b} \right) 
%
+ 4\,{{\xi}^2}\,\left( G{}\!\,_{;}{}_{c'}{}_{d'}\,G{}\!\,_{;}{}_{a}{}_{b} + 
    G\,G{}\!\,_{;}{}_{a}{}_{b}{}_{c'}{}_{d'} \right)  \cr
%
&& -2\,\xi\,\left(1-2\,\xi\right)\,
  \left( G{}\!\,_{;}{}_{b}\,G{}\!\,_{;}{}_{c'}{}_{a}{}_{d'} + 
    G{}\!\,_{;}{}_{a}\,G{}\!\,_{;}{}_{c'}{}_{b}{}_{d'} + 
    G{}\!\,_{;}{}_{d'}\,G{}\!\,_{;}{}_{a}{}_{b}{}_{c'} + 
    G{}\!\,_{;}{}_{c'}\,G{}\!\,_{;}{}_{a}{}_{b}{}_{d'} \right)  \cr
%
&& + 2\,\xi\,\left(1-2\,\xi\right)\,\left( 
G{}\!\,_{;}{}_{a}\,G{}\!\,_{;}{}_{b}\,
     {R{}_{c'}{}_{d'}} + G{}\!\,_{;}{}_{c'}\,G{}\!\,_{;}{}_{d'}\,
     {R{}_{a}{}_{b}} \right)  \cr
%
&&  -4\,{{\xi}^2}\,\left( G{}\!\,_{;}{}_{a}{}_{b}\,{R{}_{c'}{}_{d'}} + 
    G{}\!\,_{;}{}_{c'}{}_{d'}\,{R{}_{a}{}_{b}} \right)  G
%
 +  2\,{{\xi}^2}\,{R{}_{c'}{}_{d'}}\,{R{}_{a}{}_{b}} {G^2}
\end{eqnarray}
\begin{eqnarray}
8 \tilde N'_{ab} \left[ G \right]&=& 
%
2\,\left(1-2\,\xi\right) \left(
   \left(2\,\xi-{\frac{1}{2}}\right)\,G{}\!\,_{;}{}_{p'}{}_{b}\,
  G{}\!\,_{;}{}^{p'}{}_{a}
 + \xi\,\left( G{}\!\,_{;}{}_{b}\,G{}\!\,_{;}{}_{p'}{}_{a}{}^{p'} + 
    G{}\!\,_{;}{}_{a}\,G{}\!\,_{;}{}_{p'}{}_{b}{}^{p'} \right)  
\right)\cr &&
%
-4\,\xi \left(
    \left(2\,\xi-{\frac{1}{2}}\right)\,G{}\!\,_{;}{}^{p'}\,
  G{}\!\,_{;}{}_{a}{}_{b}{}_{p'}
  + \xi\,\left( G{}\!\,_{;}{}_{p'}{}^{p'}\,G{}\!\,_{;}{}_{a}{}_{b} + 
    G\,G{}\!\,_{;}{}_{a}{}_{b}{}_{p'}{}^{p'} \right) 
\right) \cr
%
&& - \left({m^2}+\xi R'\right)\,\left( 
\left(1-2\,\xi\right)\,G{}\!\,_{;}{}_{a}\,
     G{}\!\,_{;}{}_{b} - 2\,G\,\xi\,G{}\!\,_{;}{}_{a}{}_{b} \right)  \cr
%
&& + 2\,\xi\,\left( \left(2\,\xi-{\frac{1}{2}}\right)\,G{}\!\,_{;}{}_{p'}\,
     G{}\!\,_{;}{}^{p'} + 2\,G\,\xi\,G{}\!\,_{;}{}_{p'}{}^{p'} \right) \,
  {R{}_{a}{}_{b}} \cr
%
&& - \left({m^2}+\xi R'\right)\,\xi\,{R{}_{a}{}_{b}} {G^2}
\end{eqnarray}
\begin{eqnarray}
8 \tilde N \left[ G \right]&=& 
2\,{{\left(2\,\xi-{\frac{1}{2}}\right)}^2}\,G{}\!\,_{;}{}_{p'}{}_{q}\,
  G{}\!\,_{;}{}^{p'}{}^{q}
+ 4\,{{\xi}^2}\,\left( G{}\!\,_{;}{}_{p'}{}^{p'}\,G{}\!\,_{;}{}_{q}{}^{q} + 
    G\,G{}\!\,_{;}{}_{p}{}^{p}{}_{q'}{}^{q'} \right)  \cr
&& + 4\,\xi\,\left(2\,\xi-{\frac{1}{2}}\right)\,
  \left( G{}\!\,_{;}{}_{p}\,G{}\!\,_{;}{}_{q'}{}^{p}{}^{q'} + 
    G{}\!\,_{;}{}^{p'}\,G{}\!\,_{;}{}_{q}{}^{q}{}_{p'} \right)  \cr
&& - \left(2\,\xi-{\frac{1}{2}}\right)\,
  \left( \left({m^2}+\xi R\right)\,G{}\!\,_{;}{}_{p'}\,G{}\!\,_{;}{}^{p'} + 
    \left({m^2}+\xi R'\right)\,G{}\!\,_{;}{}_{p}\,G{}\!\,^{;}{}^{p} \right)  \cr
&& - 2\,\xi\,\left( \left({m^2}+\xi R\right)\,G{}\!\,_{;}{}_{p'}{}^{p'} + 
    \left({m^2}+\xi R'\right)\,G{}\!\,_{;}{}_{p}{}^{p} \right)  G \cr
&& {\frac{1}{2}} \left({m^2}+\xi R\right)\,\left({m^2}+\xi R'\right) {G^2}
\end{eqnarray}
\end{mathletters}

%
%
For a massless, conformally coupled scalar field
($\xi=\frac{1}{6}$ and  $m=0$), the noise kernel functional is
\begin{mathletters}
\begin{eqnarray}
72  \tilde N_{abc'd'} \left[ G \right]&=& 
%
4\,\left( G{}\!\,_{;}{}_{c'}{}_{b}\,G{}\!\,_{;}{}_{d'}{}_{a} + 
    G{}\!\,_{;}{}_{c'}{}_{a}\,G{}\!\,_{;}{}_{d'}{}_{b} \right) 
%
+ G{}\!\,_{;}{}_{c'}{}_{d'}\,G{}\!\,_{;}{}_{a}{}_{b} + 
  G\,G{}\!\,_{;}{}_{a}{}_{b}{}_{c'}{}_{d'} \cr
%
&& -2\,\left( G{}\!\,_{;}{}_{b}\,G{}\!\,_{;}{}_{c'}{}_{a}{}_{d'} + 
    G{}\!\,_{;}{}_{a}\,G{}\!\,_{;}{}_{c'}{}_{b}{}_{d'} + 
    G{}\!\,_{;}{}_{d'}\,G{}\!\,_{;}{}_{a}{}_{b}{}_{c'} + 
    G{}\!\,_{;}{}_{c'}\,G{}\!\,_{;}{}_{a}{}_{b}{}_{d'} \right)  \cr
%
&& + 2\,\left( G{}\!\,_{;}{}_{a}\,G{}\!\,_{;}{}_{b}\,{R{}_{c'}{}_{d'}} + 
    G{}\!\,_{;}{}_{c'}\,G{}\!\,_{;}{}_{d'}\,{R{}_{a}{}_{b}} \right)  \cr
%
&& - \left( G{}\!\,_{;}{}_{a}{}_{b}\,{R{}_{c'}{}_{d'}} + 
  G{}\!\,_{;}{}_{c'}{}_{d'}\,{R{}_{a}{}_{b}}\right) G
%
 -{\frac{1}{2}}  {R{}_{c'}{}_{d'}}\,{R{}_{a}{}_{b}} {G^2}
\end{eqnarray}
\begin{eqnarray}
 288  \tilde N'_{ab} \left[ G \right]&=& 
%
8 \left(
 -  G{}\!\,_{;}{}_{p'}{}_{b}\,G{}\!\,_{;}{}^{p'}{}_{a}
 + G{}\!\,_{;}{}_{b}\,G{}\!\,_{;}{}_{p'}{}_{a}{}^{p'} + 
  G{}\!\,_{;}{}_{a}\,G{}\!\,_{;}{}_{p'}{}_{b}{}^{p'} 
\right)\cr &&
%
4 \left(
    G{}\!\,_{;}{}^{p'}\,G{}\!\,_{;}{}_{a}{}_{b}{}_{p'}
  - G{}\!\,_{;}{}_{p'}{}^{p'}\,G{}\!\,_{;}{}_{a}{}_{b} + 
  G\,G{}\!\,_{;}{}_{a}{}_{b}{}_{p'}{}^{p'}
\right) \cr
%
&& - 2\,{R'}\,\left( 2\,G{}\!\,_{;}{}_{a}\,G{}\!\,_{;}{}_{b} - 
    G\,G{}\!\,_{;}{}_{a}{}_{b} \right)  \cr
%
&&  -2\,\left( G{}\!\,_{;}{}_{p'}\,G{}\!\,_{;}{}^{p'} - 
    2\,G\,G{}\!\,_{;}{}_{p'}{}^{p'} \right) \,{R{}_{a}{}_{b}} 
%
 - {R'}\,{R{}_{a}{}_{b}} {G^2}
\end{eqnarray}
\begin{eqnarray}
 288 \tilde N \left[ G \right]&=& 
2\,G{}\!\,_{;}{}_{p'}{}_{q}\,G{}\!\,_{;}{}^{p'}{}^{q}
+ 4\,\left( G{}\!\,_{;}{}_{p'}{}^{p'}\,G{}\!\,_{;}{}_{q}{}^{q} + 
    G\,G{}\!\,_{;}{}_{p}{}^{p}{}_{q'}{}^{q'} \right)  \cr
&& - 4\,\left( G{}\!\,_{;}{}_{p}\,G{}\!\,_{;}{}_{q'}{}^{p}{}^{q'} + 
    G{}\!\,_{;}{}^{p'}\,G{}\!\,_{;}{}_{q}{}^{q}{}_{p'} \right)  \cr
&& + R\,G{}\!\,_{;}{}_{p'}\,G{}\!\,_{;}{}^{p'} + 
  {R'}\,G{}\!\,_{;}{}_{p}\,G{}\!\,^{;}{}^{p} \cr
&& - 2\,\left( R\,G{}\!\,_{;}{}_{p'}{}^{p'} + {R'}\,G{}\!\,_{;}{}_{p}{}^{p} 
\right)
     G 
+  {\frac{1}{2}} R\,{R'} {G^2}
\end{eqnarray}
\end{mathletters}

%
%
For the minimal coupling ($\xi=0$) case:
\begin{mathletters}
\begin{equation}
8  \tilde N_{abc'd'} \left[ G \right] = 
%
G{}\!\,_{;}{}_{c'}{}_{b}\,G{}\!\,_{;}{}_{d'}{}_{a} + 
  G{}\!\,_{;}{}_{c'}{}_{a}\,G{}\!\,_{;}{}_{d'}{}_{b}
\end{equation}
\begin{equation}
8 \tilde N'_{ab} \left[ G \right] = 
 - G{}\!\,_{;}{}_{p'}{}_{b}\,G{}\!\,_{;}{}^{p'}{}_{a}
 - {m^2}\,G{}\!\,_{;}{}_{a}\,G{}\!\,_{;}{}_{b}
\end{equation}
\begin{eqnarray}
8  \tilde N \left[ G \right]&=& 
{\frac{1}{2}} \left(
   G{}\!\,_{;}{}_{p'}{}_{q}\,G{}\!\,_{;}{}^{p'}{}^{q}
 + {m^2}\,\left( G{}\!\,_{;}{}_{p'}\,G{}\!\,_{;}{}^{p'} + 
    G{}\!\,_{;}{}_{p}\,G{}\!\,^{;}{}^{p} \right) 
 + {m^4} {G^2}
\right)
\end{eqnarray}
\end{mathletters}

\subsection{Trace of the Noise Kernel}
\label{sec-ps-trace}

One of the most interesting and surprising results to come  out of the
investigations undertaken in the 1970's of the quantum stress
tensor was the discovery of the trace anomaly\cite{Duff74}. When the trace of
the stress tensor $T=g^{ab}T_{ab}$ is evaluated for a field
configuration that satisties the field equation
\begin{equation}
\frac{\delta S[\phi]}{\delta \phi(x)} = 0 \Rightarrow
\left( \Box - \xi R - m^2\right) \phi  = 0,
\label{ref-fieldeqn}
\end{equation}
the trace is seen to vanish for the massless conformally coupled case.
When this analysis is carried over to the renormalized expectation
value of the quantum stress tensor, the trace no longer vanishes.
Wald \cite{Wald78}
showed this was due to the failure of the renormalized Hadamard function
$G_{\rm ren}(x,x')$ to be symmetric in $x$ and $x'$, implying it
does not necessarily
satisfy the field equation (\ref{ref-fieldeqn}) in the
variable $x'$. (We discuss in the next section the definition of 
$G_{\rm ren}(x,x')$ in the context of point separation.)

With this in  mind, we can now determine the noise associated with the
trace. Taking the trace at both points $x$ and $y$ of the noise kernel
functional (\ref{noiker}):
\begin{eqnarray}
N\left[ G \right] &=& g^{ab}\,g^{c'd'}\, N_{abc'd'}\left[ G \right] \cr
&=&
- 3\,G\,\xi
    \left\{
        \left({m^2} + {1\over 2} \xi R \right) \,{G{}_;{}_{p'}{}^{p'}} 
      + \left({m^2} + {1\over 2} \xi R'\right) \,{G{}_;{}_{p}{}^{p}} 
    \right\} \cr
&&
   + {{9\,{{\xi }^2}}\over 2} 
       \left\{
           {G{}_;{}_{p'}{}^{p'}}\,{G{}_;{}_{p}{}^{p}}
          + G\,{G{}_;{}_{p}{}^{p}{}_{p'}{}^{p'}} 
       \right\}
    +\left({m^2} + {1\over 2} \xi R \right)  \,
     \left({m^2} + {1\over 2} \xi R'\right)   G^2
\cr
&&+ 3 \left( {1\over 6} - \xi  \right) 
      \left\{
        +3 {{\left( {1\over 6} - \xi  \right) }} 
              {G{}_;{}_{p'}{}_{p}}\,{G{}_;{}^{p'}{}^{p}}  
        -3\xi
           \left(
                {G{}_;{}_{p}}\,{G{}_;{}_{p'}{}^{p}{}^{p'}}
             +  {G{}_;{}_{p'}}\,{G{}_;{}_{p}{}^{p}{}^{p'}} 
           \right)
      \right.\cr
&&\left.\hspace{25mm}
          \left({m^2} + {1\over 2} \xi R \right) 
\,{G{}_;{}_{p'}}\,{G{}_;{}^{p'}}
        + \left({m^2} + {1\over 2} \xi R'\right) \,{G{}_;{}_{p}}\,{G{}^;{}^{p}}
\right\}
\end{eqnarray}
For the massless conformal case, this reduces to
\begin{equation}
N\left[ G \right] = \frac{1}{144}\left\{
R R' G^2 - 6G\left(R \Box' + R' \Box\right) G
  + 18\left( \left(\Box G\right)\left(\Box' G\right)+ \Box' \Box  G\right)
\right\}
\end{equation}
which holds for any function $G(x,y)$. For  $G$ being the Green function, 
it satisfies the field equation (\ref{ref-fieldeqn}):
\begin{equation}
\Box G = (m^2 + \xi R) G
\end{equation}
We will only assume the Green function satisfies the field equation
in its first variable. Using the fact $\Box' R=0$ (because the covariant 
derivatives
act at a different point than at which $R$ is supported), it follows that
\begin{equation}
\Box' \Box  G = (m^2 + \xi R)\Box' G.
\end{equation}
With these results, the noise kernel trace becomes
\begin{eqnarray}
N\left[ G \right] &=& \frac{1}{2} 
\left(
      {m^2}\,\left( 1 - 3\,\xi  \right)
    + 3\,R\,\left( {1\over 6} - \xi  \right) \,\xi  
\right) \cr
&&\hspace{30mm}\times
  \left\{
       {G^2}\,\left( 2\,{m^2} + {R'} \,\xi  \right)
            + \left( 1 - 6\,\xi  \right) \,{G{}_;{}_{p'}}\,{G{}_;{}^{p'}}
            - 6\,G\,\xi \,{G{}_;{}_{p'}{}^{p'}} 
  \right\} \cr
&&+ \frac{1}{2} \left( {1\over 6} - \xi  \right) 
\left\{
     3\,\left( 2\,{m^2} + {R'}\,\xi  \right) \,{G{}_;{}_{p}}\,{G{}^;{}^{p}}
  - 18\,\xi \,{G{}_;{}_{p}}\,{G{}_;{}_{p'}{}^{p}{}^{p'}} 
\right.\cr&&\hspace{30mm}\left.
  + 18\,\left( {1\over 6} - \xi  \right) \,
         {G{}_;{}_{p'}{}_{p}}\,{G{}_;{}^{p'}{}^{p}}  
\right\}
\end{eqnarray}
It vanishes for the massless conformal case.  We have thus shown,
based solely on the definition of the point separated noise kernel, there
is no noise associated with the trace anomaly. 
Our result is completely general since we have assumed the Green function 
is only satisfying the field equations in its first variable. This condition 
holds not just for the classical field case, but also for the regularized
quantum case, where we do not expect the Green function to satisfy the
field equation in both variables. One can see this result from a simple 
observation: Since the trace anomaly is known to be locally determined
and quantum state independent, whereas  the noise present in the quantum field 
is
non-local, it is hard to find a noise associated with it. 
This is in agreement with previous findings \cite{CH94,HuSin,CV96},
derived from the Feynman-Vernon influence functional formalism \cite{if}.


\section{Regularization of the Noise Kernel}
\label{sec-ps-reg}


We pointed out earlier that field quantities defined at two
separated points  possess important information which could be the
starting point for probes into possible extended structures of spacetime.
Moving in the other (homeward) direction, it is of interest to see how  
fluctuations of energy
momentum (loosely, noise) would enter in the ordinary (point-wise) quantum field 
theory in helping
us to address a new set of issues  such as a) whether the fluctuations to mean 
ratio can act as a 
criterion for the validity of semiclassical (note it is not that simple, see 
\cite{PH00}), 
b) whether the fluctuations in the vacuum energy density which drives  
inflationary 
violates the positive energy condition,  c) deriving  structure formation from 
quantum fluctuations, or 
d) deriving  general relativity as a low energy effective theory in the 
geometro- hydrodynamic limit \cite{stogra}. 
For these inquires we need to examine the consequences of taking or reaching  
the coincident limit 
and to construct regularization procedures to treat the ultraviolet divergences 
in order to obtain a finite result for the noise kernel in this limit.

We can see from the point separated form of the stress tensor
(\ref{ref-emt-PSdefine}) what we need to regularize is the
Green function $G^{(1)}(x,x')$. Once the Green function has been
regularized such that it is smooth and has a well defined
$x'\rightarrow x$ limit, the stress tensor will be well defined.
In Minkowski space, this issue is easily resolved by a ''normal
ordering'' prescription, which hinges on the existence of a unique
vacuum.
Unfortunately, for a general curved spacetime, there is no unique
vacuum, and hence, no unique mode expansion on which to build a 
normal ordering prescription. But we can still ask if there is a
way to determine a contribution we can subtract to yield a unique
quantum stress tensor.
Here we follow the prescription of Wald \cite{Wald75},
and Adler {\it et. al.} \cite{ALN77} (with corrections \cite{Wald78})
summarized in \cite{Wald94}. We give a short synopsis below
as it will be referred to in subsequent discussions and later papers.

The idea builds on the
fact that for 
$G(x,x')_\omega = \langle \omega |\hat\phi(x)\hat\phi(x')|\omega \rangle$,
the function
\begin{equation}
F(x,x') = G(x,x')_{\omega_1} - G(x,x')_{\omega_2}
\end{equation}
is a smooth function of $x$ and $x'$,
where $\omega_1$ and $\omega_2$ denote two different states.
 This means the difference
between the stress tensor for two states is well defined for the
point separation scheme, {\it i.e.},
\begin{equation}
{\cal F}_{ab} = \frac{1}{2}\lim_{x'\rightarrow x} {\cal T}_{ab}
  \left(F(x,x') + F(x',x) \right)
\label{ref-reg-difftab}
\end{equation}
is well defined. So  a bi-distribution $G^L(x,x')$ might be useful  for the 
vacuum subtraction. 
At first, it would seem unlikely we could find such a unique 
bi-distribution. Wald found that with the introduction of four axioms for
the regularized stress tensor
\begin{equation}
\langle \hat T_{ab}(x) \rangle_{\rm ren} = \lim_{x'\rightarrow x}
\frac{1}{2} {\cal T}_{ab}\left(
{G^{(1)}}(x,x') - {G^{L}}(x,x') \right)
\label{ref-eq-STren-define}
\end{equation}
 $G^L(x,x')$ is uniquely determined,
up to a local conserved curvature term. The Wald axioms are 
\cite{Wald78,Wald94}:
\begin{enumerate}
  \item The difference between the stress tensor for two states should
        agree with (\ref{ref-reg-difftab});
  \item The stress tensor should be local with respect to the state of
        the field;
  \item For all states, the stress tensor is conserved:
                    $\nabla^a \langle T_{ab}\rangle = 0$;
  \item In Minkowski space, the result 
               $\langle 0| T_{ab}|0\rangle = 0$ is recovered.
\end{enumerate}

We are still left with the problem of determining the form of such a
bi-distribution.  Hadamard \cite{Hadamard} showed that the Green function for 
a large class of states takes the form (in four spacetime dimensions)
\begin{equation}
G^L(x,x') = \frac{1}{8\pi^2}\left(\frac{2U(x,x')}{\sigma}
  +V(x,x')\log\sigma + W(x,x') \right)
\label{ref-HadamardAnsatz}
\end{equation}
with $U(x,x')$, $V(x,x')$ and $W(x,x')$ being smooth functions. 
\footnote{When working in the
Lorentz sector of a field theory,
{\it i.e.}, when the metric signature is $(-,+,+,+)$, as opposed to the
Euclidean sector with the signature $(+,+,+,+)$,
we must modify the above function to
account for null geodesics, since $\sigma(x,x')=0$ for null separated
$x$ and $x'$. This problem can be overcome by using $\sigma \rightarrow \sigma +
2i\epsilon (t-t') + \epsilon^2$. Here, we will work only with
geometries that possess Euclidean sectors and carry out our analysis
with Riemannian geometries and only at the end continue back to the
Lorentzian geometry. Nonetheless, this presents no difficulty. At any
point in the analysis the above replacement for $\sigma$ can be performed.}
We refer to Eqn (\ref{ref-HadamardAnsatz})
as the ``Hadamard ansatz''.

Since the functions $V(x,x')$ and $W(x,x')$ are smooth functions, they
can be expanded as
\begin{mathletters}
\label{ref-vwseries}
\begin{eqnarray}
V(x,x') = \sum_{n=0}^\infty v_n(x,x') \sigma^n \\
W(x,x') = \sum_{n=0}^\infty w_n(x,x') \sigma^n
\end{eqnarray}
\end{mathletters}
with the $v_n$'s and $w_n$'s themselves smooth functions. These functions
and $U(x,x')$ are determined by substituting $G^L(x,x')$ in the wave equation
$K G^L(x,x') = 0$ and equating to zero the coefficients of the explicitly
appearing powers of $\sigma^n$ and $\sigma^n\log\sigma$. Doing so, we 
get the infinite set of equations
\begin{mathletters}
\begin{eqnarray}
                              U(x,x') &=& {\Delta\!^{1/2}}; \\
                  2 H_0 v_0 + K {\Delta\!^{1/2}} &=& 0; \label{ref-v0eqn}\\
               2n H_n v_n + K v_{n-1} &=& 0, \quad n \ge 1;\label{ref-vneqn}\\
 2H_{2n} v_n + 2n H_n w_n + K w_{n-1} &=& 0, \quad n \ge 1\label{ref-wneqn}
\label{ref-vweqns}
\end{eqnarray}
with
\begin{equation}
H_n = \sigma^{;p} \nabla_p + \left(n-1 
    +\frac{1}{2}\left(\Box\sigma\right)\right)
\end{equation}
\end{mathletters}
>From Eqs (\ref{ref-vweqns}), the functions $v_n$ are completely determined.
In fact, they are symmetric functions, and hence $V(x,x')$ is a symmetric
function of $x$ and $x'$. On the other hand, the field equations only determine
$w_n$, $n \ge 1$, leaving $w_0(x,x')$ completely  arbitrary. This reflects the
state dependence of the Hadamard form above. Moreover, even if $w_0(x,x')$ is
chosen to be symmetric, this does not guarantee that $W(x,x')$ will be. By
using axiom (4)  $w_0(x,x') \equiv 0$. With this choice, the Minkowski
spacetime
limit is
\begin{equation}
G^L = \frac{1}{(2\pi)^2} \frac{1}{\sigma}
\end{equation}
where now $2 \sigma =  (t-t')^2 - ({\bf x}-{\bf x}')^2$ and this
corresponds to the correct vacuum contribution that needs to be subtracted.

With this choice though, we are left with a $G^L(x,x')$ which is not symmetric
and hence does not satisfy the field equation at $x'$, for fixed $x$. Wald
\cite{Wald78} showed this in turn implies axiom (3) is not satisfied. He
resolved this problem by adding to the regularized stress tensor a term which 
cancels that which breaks the conservation of the old stress tensor:
\begin{equation}
\langle T^{\rm new}_{ab}\rangle = \langle T^{\rm old}_{ab}\rangle +
\frac{1}{2(4\pi)^2} g_{ab} \left[ v_1 \right]
\end{equation}
where $\left[ v_1 \right] = v_1(x,x)$ is the coincident limit of the
$n=1$ solution of Eq (\ref{ref-vweqns}). This yields a stress tensor
which satisfies all four axioms and produces the well known trace anomaly
$\langle T_{a}{}^a\rangle = \left[ v_1 \right]/8\pi^2$. We can view this
redefinition as taking place at the level of the stress tensor operator
via
\begin{equation}
\hat T_{ab} \rightarrow \hat T_{ab} 
+ \frac{\hat 1}{2(4\pi)^2} g_{ab} \left[ v_1 \right]
\end{equation}
Since this amounts to a constant shift of the stress tensor operator, it
will have no effect on the noise kernel or fluctuations, as they are
 the variance about the mean. This is further supported by
the fact that there is no noise associated with the trace. Since
this result was derived by only assuming that the Green function satisfies the
field equation in {\em one} of its variables, it is independent of the
issue of the lack of symmetry in the Hadamard ansatz 
(\ref{ref-HadamardAnsatz}).

Using the above formalism we now derive  the coincident limit expression for 
the noise kernel
(\ref{ref-noise-kernel}).
To get a meaningful result, we must work with the regularization of the 
Wightman function, obtained by following the same procedure outlined
above for the Hadamard function:
\begin{equation}
G_{\rm ren}(x,y) \equiv G_{\rm ren,+}(x,y) =
	G_+(x,y) - G^L(x,y)
\end{equation}
In doing this, we assume the singular structure of the Wightman function
is the same as that for the Hadamard function. In all applications, this
is indeed the case. Moreover, when we compute the coincident limit
of $N_{abc'd'}$, we will be working in the Euclidean section where there is no
issue of operator ordering.   For now we only consider spacetimes
with no time dependence present in the final coincident limit result, 
so there is also no issue of Wick rotation back to a Minkowski signature.
If this was the case, then care must be taken as to
whether we are considering
$\left[ N_{abc'd'} \left[ G_{\rm ren,+}(x,y) \right] \right]$ or
$\left[ N_{abc'd'} \left[ G_{\rm ren,+}(y,x) \right] \right]$.

We now have all the information we need to compute the coincident limit
of the noise kernel (\ref{ref-noise-kernel}). Since the point
separated noise kernel $N_{abc'd'}(x,y)$ involves covariant derivatives at 
the two points at which it has support, when we take the coincident limit
we can use Synge's theorem (\ref{ref-Synge's}) to move the derivatives
acting at $y$ to ones acting at $x$. Due to the long length of the noise
kernel expression, we will only  give an example by examining a single
term. 

Consider a typical term from the noise kernel functional
(\ref{ref-noise-kernel}):
\begin{equation}
G_{{\rm ren}}{}_{;}{}_{c'}{}_{b}\,G_{{\rm ren}}{}_{;}{}_{d'}{}_{a} + G_{{\rm 
ren}}{}_{;}{}_{c'}{}_{a}\,G_{{\rm ren}}{}_{;}{}_{d'}{}_{b}
\end{equation}
Recall the noise kernel itself is related to the noise kernel functional
via
\begin{equation}
N_{abc'd'} = N_{abc'd'}\left[G_{ren}(x,y)\right]
                + N_{abc'd'}\left[G_{ren}(y,x)\right].
\end{equation}
This is implemented on our typical term by adding to it the same term,
but now with the roles of $x$ and $y$ reversed, so we have to consider
\begin{equation}
G_{{\rm ren}}{}_{;}{}_{c'}{}_{b}\,G_{{\rm ren}}{}_{;}{}_{d'}{}_{a} + G_{{\rm 
ren}}{}_{;}{}_{a'}{}_{d}\,G_{{\rm ren}}{}_{;}{}_{b'}{}_{c} + G_{{\rm 
ren}}{}_{;}{}_{c'}{}_{a}\,G_{{\rm ren}}{}_{;}{}_{d'}{}_{b} + G_{{\rm 
ren}}{}_{;}{}_{a'}{}_{c}\,G_{{\rm ren}}{}_{;}{}_{b'}{}_{d}
\end{equation}
It is to this form that we can take the coincident limit:
\begin{equation}
{\left[G_{{\rm ren}}{}_{;}{}_{c'}{}_{b}\right]}\,
  {\left[G_{{\rm ren}}{}_{;}{}_{d'}{}_{a}\right]} + {\left[G_{{\rm 
ren}}{}_{;}{}_{a'}{}_{d}\right]}\,
  {\left[G_{{\rm ren}}{}_{;}{}_{b'}{}_{c}\right]} + {\left[G_{{\rm 
ren}}{}_{;}{}_{c'}{}_{a}\right]}\,
  {\left[G_{{\rm ren}}{}_{;}{}_{d'}{}_{b}\right]} + {\left[G_{{\rm 
ren}}{}_{;}{}_{a'}{}_{c}\right]}\,
  {\left[G_{{\rm ren}}{}_{;}{}_{b'}{}_{d}\right]}
\end{equation}
We can now apply Synge's theorem:
\begin{eqnarray}
&&\hspace{3mm}
\left( {\left[G_{{\rm ren}}{}_{;}{}_{a}\right]}{}_{;}{}_{d} - 
    {\left[G_{{\rm ren}}{}_{;}{}_{a}{}_{d}\right]} \right) \,
  \left( {\left[G_{{\rm ren}}{}_{;}{}_{b}\right]}{}_{;}{}_{c} - 
    {\left[G_{{\rm ren}}{}_{;}{}_{b}{}_{c}\right]} \right)  \cr
&& + \left( {\left[G_{{\rm ren}}{}_{;}{}_{d}\right]}{}_{;}{}_{a} - 
    {\left[G_{{\rm ren}}{}_{;}{}_{a}{}_{d}\right]} \right) \,
  \left( {\left[G_{{\rm ren}}{}_{;}{}_{c}\right]}{}_{;}{}_{b} - 
    {\left[G_{{\rm ren}}{}_{;}{}_{b}{}_{c}\right]} \right)  \cr
&& + \left( {\left[G_{{\rm ren}}{}_{;}{}_{a}\right]}{}_{;}{}_{c} - 
    {\left[G_{{\rm ren}}{}_{;}{}_{a}{}_{c}\right]} \right) \,
  \left( {\left[G_{{\rm ren}}{}_{;}{}_{b}\right]}{}_{;}{}_{d} - 
    {\left[G_{{\rm ren}}{}_{;}{}_{b}{}_{d}\right]} \right)  \cr
&& + \left( {\left[G_{{\rm ren}}{}_{;}{}_{c}\right]}{}_{;}{}_{a} - 
    {\left[G_{{\rm ren}}{}_{;}{}_{a}{}_{c}\right]} \right) \,
  \left( {\left[G_{{\rm ren}}{}_{;}{}_{d}\right]}{}_{;}{}_{b} - 
    {\left[G_{{\rm ren}}{}_{;}{}_{b}{}_{d}\right]} \right).
\end{eqnarray}
This is the desired form for once we have an end point expansion
of $G_{{\rm ren}}$, it will be straightforward to compute the above
expression. We will discuss the details of such an evaluation in
the context of symbolic computations in a companion paper in this series.

The final result for the coincident limit of the noise kernel is
broken down into a rank four and rank two tensor and a scalar
according to
%
%
%
\begin{mathletters}
\label{coincident-noise}
\begin{equation}
 \left[ N_{abc'd'} \right]  = 
    N'_{abcd}
  + g_{ab} N''_{cd}
 + g_{cd} N''_{ab}
 + g_{ab}g_{cd} N'''
\end{equation}
The  two tensors and the scalar have the following form in terms of the
concident limit and derivatives of the renormalized Green function and
its derivatives:
\begin{eqnarray}
8 N_{abcd} &=& 
%
{{\left(1-2\,\xi\right)}^2} \left\{
   \left( {\left[G_{ren}{}\!\,_{;}{}_{a}\right]}{}\!\,_{;}{}_{c} - 
    {\left[G_{ren}{}\!\,_{;}{}_{a}{}_{c}\right]} \right) \,
  \left( {\left[G_{ren}{}\!\,_{;}{}_{b}\right]}{}\!\,_{;}{}_{d} - 
    {\left[G_{ren}{}\!\,_{;}{}_{b}{}_{d}\right]} \right)  \right.  \cr
&&\hspace{16mm} +  \left( 
{\left[G_{ren}{}\!\,_{;}{}_{c}\right]}{}\!\,_{;}{}_{a} - 
    {\left[G_{ren}{}\!\,_{;}{}_{a}{}_{c}\right]} \right) \,
  \left( {\left[G_{ren}{}\!\,_{;}{}_{d}\right]}{}\!\,_{;}{}_{b} - 
    {\left[G_{ren}{}\!\,_{;}{}_{b}{}_{d}\right]} \right)    \cr
&&\hspace{16mm} +  \left( 
{\left[G_{ren}{}\!\,_{;}{}_{a}\right]}{}\!\,_{;}{}_{d} - 
    {\left[G_{ren}{}\!\,_{;}{}_{a}{}_{d}\right]} \right) \,
  \left( {\left[G_{ren}{}\!\,_{;}{}_{b}\right]}{}\!\,_{;}{}_{c} - 
    {\left[G_{ren}{}\!\,_{;}{}_{b}{}_{c}\right]} \right)    \cr && \left.
  \hspace{16mm} + \left( 
{\left[G_{ren}{}\!\,_{;}{}_{d}\right]}{}\!\,_{;}{}_{a} - 
    {\left[G_{ren}{}\!\,_{;}{}_{a}{}_{d}\right]} \right) \,
  \left( {\left[G_{ren}{}\!\,_{;}{}_{c}\right]}{}\!\,_{;}{}_{b} - 
    {\left[G_{ren}{}\!\,_{;}{}_{b}{}_{c}\right]} \right)    \right\} \cr
%
&& + 4\,{{\xi}^2} \left\{
   \left( -{\left[G_{ren}{}\!\,_{;}{}_{a}\right]}{}\!\,_{;}{}_{b} - 
    {\left[G_{ren}{}\!\,_{;}{}_{b}\right]}{}\!\,_{;}{}_{a} + 
    {\left[G_{ren}\right]}{}\!\,_{;}{}_{a}{}_{b} + 
    {\left[G_{ren}{}\!\,_{;}{}_{a}{}_{b}\right]} \right) \,
  {\left[G_{ren}{}\!\,_{;}{}_{c}{}_{d}\right]} \right.  \cr
&&\hspace{14mm} +  {\left[G_{ren}{}\!\,_{;}{}_{a}{}_{b}\right]}\,
  \left( -{\left[G_{ren}{}\!\,_{;}{}_{c}\right]}{}\!\,_{;}{}_{d} - 
    {\left[G_{ren}{}\!\,_{;}{}_{d}\right]}{}\!\,_{;}{}_{c} + 
    {\left[G_{ren}\right]}{}\!\,_{;}{}_{c}{}_{d} + 
    {\left[G_{ren}{}\!\,_{;}{}_{c}{}_{d}\right]} \right)    \cr
&&\hspace{14mm} +  {\left[G_{ren}\right]}\,\left( 
-{\left[G_{ren}{}\!\,_{;}{}_{a}{}_{b}{}_{c}
       \right]}{}\!\,_{;}{}_{d} - 
    {\left[G_{ren}{}\!\,_{;}{}_{a}{}_{b}{}_{d}\right]}{}\!\,_{;}{}_{c} + 
    {\left[G_{ren}{}\!\,_{;}{}_{a}{}_{b}\right]}{}\!\,_{;}{}_{c}{}_{d} + 
    {\left[G_{ren}{}\!\,_{;}{}_{a}{}_{b}{}_{d}{}_{c}\right]} \right)    \cr && 
\left.
  \hspace{14mm} + {\left[G_{ren}\right]}\,\left( 
-{\left[G_{ren}{}\!\,_{;}{}_{c}{}_{d}{}_{a}
       \right]}{}\!\,_{;}{}_{b} - 
    {\left[G_{ren}{}\!\,_{;}{}_{c}{}_{d}{}_{b}\right]}{}\!\,_{;}{}_{a} + 
    {\left[G_{ren}{}\!\,_{;}{}_{c}{}_{d}\right]}{}\!\,_{;}{}_{a}{}_{b} + 
    {\left[G_{ren}{}\!\,_{;}{}_{c}{}_{d}{}_{b}{}_{a}\right]} \right)    
\right\} \cr
%
&& -2\,\xi\,\left(1-2\,\xi\right) \left\{
   \left( {\left[G_{ren}\right]}{}\!\,_{;}{}_{a} - 
    {\left[G_{ren}{}\!\,_{;}{}_{a}\right]} \right) \,
  \left( {\left[G_{ren}{}\!\,_{;}{}_{c}{}_{d}\right]}{}\!\,_{;}{}_{b} - 
    {\left[G_{ren}{}\!\,_{;}{}_{c}{}_{d}{}_{b}\right]} \right)  \right.  \cr
&&\hspace{24mm} +  \left( {\left[G_{ren}\right]}{}\!\,_{;}{}_{b} - 
    {\left[G_{ren}{}\!\,_{;}{}_{b}\right]} \right) \,
  \left( {\left[G_{ren}{}\!\,_{;}{}_{c}{}_{d}\right]}{}\!\,_{;}{}_{a} - 
    {\left[G_{ren}{}\!\,_{;}{}_{c}{}_{d}{}_{a}\right]} \right)    \cr
&&\hspace{24mm} +  \left( {\left[G_{ren}\right]}{}\!\,_{;}{}_{c} - 
    {\left[G_{ren}{}\!\,_{;}{}_{c}\right]} \right) \,
  \left( {\left[G_{ren}{}\!\,_{;}{}_{a}{}_{b}\right]}{}\!\,_{;}{}_{d} - 
    {\left[G_{ren}{}\!\,_{;}{}_{a}{}_{b}{}_{d}\right]} \right)    \cr
&&\hspace{24mm} +  \left( {\left[G_{ren}\right]}{}\!\,_{;}{}_{d} - 
    {\left[G_{ren}{}\!\,_{;}{}_{d}\right]} \right) \,
  \left( {\left[G_{ren}{}\!\,_{;}{}_{a}{}_{b}\right]}{}\!\,_{;}{}_{c} - 
    {\left[G_{ren}{}\!\,_{;}{}_{a}{}_{b}{}_{c}\right]} \right)    \cr
&&\hspace{17mm} +  {\left[G_{ren}{}\!\,_{;}{}_{a}\right]}\,
  \left( -{\left[G_{ren}{}\!\,_{;}{}_{b}{}_{c}\right]}{}\!\,_{;}{}_{d} - 
    {\left[G_{ren}{}\!\,_{;}{}_{b}{}_{d}\right]}{}\!\,_{;}{}_{c} + 
    {\left[G_{ren}{}\!\,_{;}{}_{b}\right]}{}\!\,_{;}{}_{c}{}_{d} + 
    {\left[G_{ren}{}\!\,_{;}{}_{b}{}_{d}{}_{c}\right]} \right)    \cr
&&\hspace{17mm} +  {\left[G_{ren}{}\!\,_{;}{}_{b}\right]}\,
  \left( -{\left[G_{ren}{}\!\,_{;}{}_{a}{}_{c}\right]}{}\!\,_{;}{}_{d} - 
    {\left[G_{ren}{}\!\,_{;}{}_{a}{}_{d}\right]}{}\!\,_{;}{}_{c} + 
    {\left[G_{ren}{}\!\,_{;}{}_{a}\right]}{}\!\,_{;}{}_{c}{}_{d} + 
    {\left[G_{ren}{}\!\,_{;}{}_{a}{}_{d}{}_{c}\right]} \right)    \cr
&&\hspace{17mm} +  {\left[G_{ren}{}\!\,_{;}{}_{c}\right]}\,
  \left( -{\left[G_{ren}{}\!\,_{;}{}_{a}{}_{d}\right]}{}\!\,_{;}{}_{b} - 
    {\left[G_{ren}{}\!\,_{;}{}_{b}{}_{d}\right]}{}\!\,_{;}{}_{a} + 
    {\left[G_{ren}{}\!\,_{;}{}_{d}\right]}{}\!\,_{;}{}_{a}{}_{b} + 
    {\left[G_{ren}{}\!\,_{;}{}_{b}{}_{d}{}_{a}\right]} \right)    \cr && \left.
  \hspace{17mm} + {\left[G_{ren}{}\!\,_{;}{}_{d}\right]}\,
  \left( -{\left[G_{ren}{}\!\,_{;}{}_{a}{}_{c}\right]}{}\!\,_{;}{}_{b} - 
    {\left[G_{ren}{}\!\,_{;}{}_{b}{}_{c}\right]}{}\!\,_{;}{}_{a} + 
    {\left[G_{ren}{}\!\,_{;}{}_{c}\right]}{}\!\,_{;}{}_{a}{}_{b} + 
    {\left[G_{ren}{}\!\,_{;}{}_{b}{}_{c}{}_{a}\right]} \right)    \right\} \cr
%
&& + 2\,\xi\,\left(1-2\,\xi\right) \left\{
 \left( \left( {\left[G_{ren}\right]}{}\!\,_{;}{}_{a} - 
       {\left[G_{ren}{}\!\,_{;}{}_{a}\right]} \right) \,
     \left( {\left[G_{ren}\right]}{}\!\,_{;}{}_{b} - 
       {\left[G_{ren}{}\!\,_{;}{}_{b}\right]} \right)  + 
    {\left[G_{ren}{}\!\,_{;}{}_{a}\right]}\,
     {\left[G_{ren}{}\!\,_{;}{}_{b}\right]} \right) \,{R{}_{c}{}_{d}}
\right. \cr
%
&& \hspace{23mm} + \left. 
 \left( \left( {\left[G_{ren}\right]}{}\!\,_{;}{}_{c} - 
       {\left[G_{ren}{}\!\,_{;}{}_{c}\right]} \right) \,
     \left( {\left[G_{ren}\right]}{}\!\,_{;}{}_{d} - 
       {\left[G_{ren}{}\!\,_{;}{}_{d}\right]} \right)  + 
    {\left[G_{ren}{}\!\,_{;}{}_{c}\right]}\,
     {\left[G_{ren}{}\!\,_{;}{}_{d}\right]} \right) \,{R{}_{a}{}_{b}}
\right\} \cr
%
&& + 4\,{{\xi}^2} \left\{
 \left( {\left[G_{ren}{}\!\,_{;}{}_{a}\right]}{}\!\,_{;}{}_{b} + 
    {\left[G_{ren}{}\!\,_{;}{}_{b}\right]}{}\!\,_{;}{}_{a} - 
    {\left[G_{ren}\right]}{}\!\,_{;}{}_{a}{}_{b} - 
    2\,{\left[G_{ren}{}\!\,_{;}{}_{a}{}_{b}\right]} \right) \,{R{}_{c}{}_{d}}
\right. \cr
%
&& \hspace{8mm} + \left. 
 \left( {\left[G_{ren}{}\!\,_{;}{}_{c}\right]}{}\!\,_{;}{}_{d} + 
    {\left[G_{ren}{}\!\,_{;}{}_{d}\right]}{}\!\,_{;}{}_{c} - 
    {\left[G_{ren}\right]}{}\!\,_{;}{}_{c}{}_{d} - 
    2\,{\left[G_{ren}{}\!\,_{;}{}_{c}{}_{d}\right]} \right) \,{R{}_{a}{}_{b}}
\right. \cr
%
&& \hspace{8mm} + \left. 
 {R{}_{a}{}_{b}}\,{R{}_{c}{}_{d}} {\left[G_{ren}\right]} 
\right\}  {\left[G_{ren}\right]}
\end{eqnarray}
\begin{eqnarray}
8 N'_{ab} &=& 
%
2\,\left(2\,\xi-{\frac{1}{2}}\right)\,\left(1-2\,\xi\right) \left(
 \left( {\left[G_{ren}{}\!\,_{;}{}_{b}\right]}{}\!\,_{;}{}_{p} - 
    {\left[G_{ren}{}\!\,_{;}{}_{p}{}_{b}\right]} \right) \,
  \left( {\left[G_{ren}{}\!\,_{;}{}_{a}\right]}{}\!\,^{;}{}^{p} - 
    {\left[G_{ren}{}\!\,_{;}{}_{a}{}^{p}\right]} \right)  
\right. \cr && \left. \hspace{36mm} +
 \left( {\left[G_{ren}{}\!\,_{;}{}_{p}\right]}{}\!\,_{;}{}_{b} - 
    {\left[G_{ren}{}\!\,_{;}{}_{p}{}_{b}\right]} \right) \,
  \left( {\left[G_{ren}{}\!\,^{;}{}^{p}\right]}{}\!\,_{;}{}_{a} - 
    {\left[G_{ren}{}\!\,_{;}{}_{a}{}^{p}\right]} \right)  
\right) \cr
%
&&+2\,\xi\,\left(1-2\,\xi\right) \left(
 \left( {\left[G_{ren}\right]}{}\!\,_{;}{}_{a} - 
    {\left[G_{ren}{}\!\,_{;}{}_{a}\right]} \right) \,
  \left( {\left[G_{ren}{}\!\,_{;}{}_{p}{}^{p}\right]}{}\!\,_{;}{}_{b} - 
    {\left[G_{ren}{}\!\,_{;}{}_{p}{}^{p}{}_{b}\right]} \right)  
\right. \cr && \left. \hspace{22mm} +
 \left( {\left[G_{ren}\right]}{}\!\,_{;}{}_{b} - 
    {\left[G_{ren}{}\!\,_{;}{}_{b}\right]} \right) \,
  \left( {\left[G_{ren}{}\!\,_{;}{}_{p}{}^{p}\right]}{}\!\,_{;}{}_{a} - 
    {\left[G_{ren}{}\!\,_{;}{}_{p}{}^{p}{}_{a}\right]} \right)  
\right. \cr && \left. \hspace{22mm} +
 {\left[G_{ren}{}\!\,_{;}{}_{a}\right]}\,
  \left( -{\left[G_{ren}{}\!\,_{;}{}_{p}{}_{b}\right]}{}\!\,^{;}{}^{p} - 
    {\left[G_{ren}{}\!\,_{;}{}_{b}{}^{p}\right]}{}\!\,_{;}{}_{p} + 
    {\left[G_{ren}{}\!\,_{;}{}_{b}\right]}{}\!\,_{;}{}_{p}{}^{p} + 
    {\left[G_{ren}{}\!\,_{;}{}_{b}{}^{p}{}_{p}\right]} \right)  
\right. \cr && \left. \hspace{22mm} +
 {\left[G_{ren}{}\!\,_{;}{}_{b}\right]}\,
  \left( -{\left[G_{ren}{}\!\,_{;}{}_{p}{}_{a}\right]}{}\!\,^{;}{}^{p} - 
    {\left[G_{ren}{}\!\,_{;}{}_{a}{}^{p}\right]}{}\!\,_{;}{}_{p} + 
    {\left[G_{ren}{}\!\,_{;}{}_{a}\right]}{}\!\,_{;}{}_{p}{}^{p} + 
    {\left[G_{ren}{}\!\,_{;}{}_{a}{}^{p}{}_{p}\right]} \right)  
\right) \cr
&&-4\,\xi\,\left(2\,\xi-{\frac{1}{2}}\right) \left(
 {\left[G_{ren}{}\!\,^{;}{}^{p}\right]}\,
  \left( -{\left[G_{ren}{}\!\,_{;}{}_{p}{}_{a}\right]}{}\!\,_{;}{}_{b} - 
    {\left[G_{ren}{}\!\,_{;}{}_{p}{}_{b}\right]}{}\!\,_{;}{}_{a} + 
    {\left[G_{ren}{}\!\,_{;}{}_{p}\right]}{}\!\,_{;}{}_{a}{}_{b} + 
    {\left[G_{ren}{}\!\,_{;}{}_{p}{}_{b}{}_{a}\right]} \right)  
\right. \cr && \left. \hspace{24mm} +
 \left( {\left[G_{ren}\right]}{}\!\,^{;}{}^{p} - 
    {\left[G_{ren}{}\!\,^{;}{}^{p}\right]} \right) \,
  \left( {\left[G_{ren}{}\!\,_{;}{}_{a}{}_{b}\right]}{}\!\,_{;}{}_{p} - 
    {\left[G_{ren}{}\!\,_{;}{}_{a}{}_{b}{}_{p}\right]} \right)  
\right) \cr
%
&& -2\,\left({m^2}+\xi R\right)\,\xi\,{\left[G_{ren}\right]}\,
  \left( {\left[G_{ren}{}\!\,_{;}{}_{a}\right]}{}\!\,_{;}{}_{b} + 
    {\left[G_{ren}{}\!\,_{;}{}_{b}\right]}{}\!\,_{;}{}_{a} - 
    {\left[G_{ren}\right]}{}\!\,_{;}{}_{a}{}_{b} - 
    2\,{\left[G_{ren}{}\!\,_{;}{}_{a}{}_{b}\right]} \right)  \cr
%
&& + 2\,\xi\,\left(2\,\xi-{\frac{1}{2}}\right)\,
  \left( {\left[G_{ren}\right]}{}\!\,_{;}{}_{p}\,
     {\left[G_{ren}\right]}{}\!\,^{;}{}^{p} - 
    2\,{\left[G_{ren}\right]}{}\!\,_{;}{}_{p}\,
     {\left[G_{ren}{}\!\,^{;}{}^{p}\right]} + 
    2\,{\left[G_{ren}{}\!\,_{;}{}_{p}\right]}\,
     {\left[G_{ren}{}\!\,^{;}{}^{p}\right]} \right) \,{R{}_{a}{}_{b}} \cr
%
&& -4\,{{\xi}^2}\,{\left[G_{ren}\right]}\,
  \left( 2\,{\left[G_{ren}{}\!\,_{;}{}_{p}\right]}{}\!\,^{;}{}^{p} - 
    {\left[G_{ren}\right]}{}\!\,_{;}{}_{p}{}^{p} - 
    2\,{\left[G_{ren}{}\!\,_{;}{}_{p}{}^{p}\right]} \right) \,{R{}_{a}{}_{b}} 
\cr
%
&& -2\,\left({m^2}+\xi R\right)\,\xi\,{R{}_{a}{}_{b}} 
{{{\left[G_{ren}\right]}}^2}
\end{eqnarray}
\begin{eqnarray}
8 N'''&=& 
%
2\,{{\left(2\,\xi-{\frac{1}{2}}\right)}^2} \left(
 \left( {\left[G_{ren}{}\!\,_{;}{}_{p}\right]}{}\!\,_{;}{}_{q} - 
    {\left[G_{ren}{}\!\,_{;}{}_{p}{}_{q}\right]} \right) \,
  \left( {\left[G_{ren}{}\!\,^{;}{}^{p}\right]}{}\!\,^{;}{}^{q} - 
    {\left[G_{ren}{}\!\,^{;}{}^{p}{}^{q}\right]} \right) 
\right. \cr && \left. \hspace{20mm} +
 \left( {\left[G_{ren}{}\!\,_{;}{}_{q}\right]}{}\!\,_{;}{}_{p} - 
    {\left[G_{ren}{}\!\,_{;}{}_{p}{}_{q}\right]} \right) \,
  \left( {\left[G_{ren}{}\!\,^{;}{}^{q}\right]}{}\!\,^{;}{}^{p} - 
    {\left[G_{ren}{}\!\,^{;}{}^{p}{}^{q}\right]} \right) 
\right) \cr
%
&&-8\,{{\xi}^2} \left(
 \left( 2\,{\left[G_{ren}{}\!\,_{;}{}_{p}\right]}{}\!\,^{;}{}^{p} - 
    {\left[G_{ren}\right]}{}\!\,_{;}{}_{p}{}^{p} - 
    {\left[G_{ren}{}\!\,_{;}{}_{p}{}^{p}\right]} \right) \,
  {\left[G_{ren}{}\!\,_{;}{}_{q}{}^{q}\right]}
\right. \cr && \left. \hspace{16mm} 
 {\left[G_{ren}\right]}\,\left( 2\,
     {\left[G_{ren}{}\!\,_{;}{}_{p}{}^{p}{}_{q}\right]}{}\!\,^{;}{}^{q} - 
    {\left[G_{ren}{}\!\,_{;}{}_{p}{}^{p}\right]}{}\!\,_{;}{}_{q}{}^{q} - 
    {\left[G_{ren}{}\!\,_{;}{}_{p}{}^{p}{}_{q}{}^{q}\right]} \right) 
\right) \cr
%
&& -8\,\xi\,\left(2\,\xi-{\frac{1}{2}}\right) \left(
\left( {\left[G_{ren}\right]}{}\!\,_{;}{}_{p} - 
    {\left[G_{ren}{}\!\,_{;}{}_{p}\right]} \right) \,
  {\left[G_{ren}{}\!\,_{;}{}_{q}{}^{q}{}^{p}\right]}
- {\left[G_{ren}\right]}{}\!\,_{;}{}_{p}\,
  {\left[G_{ren}{}\!\,_{;}{}_{q}{}^{q}\right]}{}\!\,^{;}{}^{p}
\right. \cr && \left. \hspace{20mm} + 
 {\left[G_{ren}{}\!\,_{;}{}_{p}\right]}\,
  \left( 2\,{\left[G_{ren}{}\!\,_{;}{}_{q}{}^{p}\right]}{}\!\,^{;}{}^{q} + 
    {\left[G_{ren}{}\!\,_{;}{}_{q}{}^{q}\right]}{}\!\,^{;}{}^{p} - 
    {\left[G_{ren}{}\!\,^{;}{}^{p}\right]}{}\!\,_{;}{}_{q}{}^{q} - 
    {\left[G_{ren}{}\!\,_{;}{}_{q}{}^{p}{}^{q}\right]} \right) 
\right) \cr
%
&&    -2\,\left({m^2}+\xi R\right)\,\left(2\,\xi-{\frac{1}{2}}\right)\,
  \left( {\left[G_{ren}\right]}{}\!\,_{;}{}_{p}\,
     {\left[G_{ren}\right]}{}\!\,^{;}{}^{p} - 
    2\,{\left[G_{ren}\right]}{}\!\,_{;}{}_{p}\,
     {\left[G_{ren}{}\!\,^{;}{}^{p}\right]} + 
    2\,{\left[G_{ren}{}\!\,_{;}{}_{p}\right]}\,
     {\left[G_{ren}{}\!\,^{;}{}^{p}\right]} \right)  \cr
%
&& +  4\,\left({m^2}+\xi R\right)\,\xi\,
  \left( 2\,{\left[G_{ren}{}\!\,_{;}{}_{p}\right]}{}\!\,^{;}{}^{p} - 
    {\left[G_{ren}\right]}{}\!\,_{;}{}_{p}{}^{p} - 
    2\,{\left[G_{ren}{}\!\,_{;}{}_{p}{}^{p}\right]} \right)  
{\left[G_{ren}\right]}\cr
%
&& + {{\left({m^2}+\xi R\right)}^2} {{{\left[G_{ren}\right]}}^2}
\end{eqnarray}
\label{noise-coin-general}
\end{mathletters}

%
%
For the conformal coupling ($\xi=\frac{1}{6}$ and  $m=0$) case:
\begin{mathletters}
\begin{eqnarray}
 72 N_{abcd} &=& 
%
4 \left\{
   \left( {\left[G_{ren}{}\!\,_{;}{}_{a}\right]}{}\!\,_{;}{}_{c} - 
    {\left[G_{ren}{}\!\,_{;}{}_{a}{}_{c}\right]} \right) \,
  \left( {\left[G_{ren}{}\!\,_{;}{}_{b}\right]}{}\!\,_{;}{}_{d} - 
    {\left[G_{ren}{}\!\,_{;}{}_{b}{}_{d}\right]} \right)  \right.  \cr
&&\hspace{2mm} +  \left( 
{\left[G_{ren}{}\!\,_{;}{}_{c}\right]}{}\!\,_{;}{}_{a} - 
    {\left[G_{ren}{}\!\,_{;}{}_{a}{}_{c}\right]} \right) \,
  \left( {\left[G_{ren}{}\!\,_{;}{}_{d}\right]}{}\!\,_{;}{}_{b} - 
    {\left[G_{ren}{}\!\,_{;}{}_{b}{}_{d}\right]} \right)    \cr
&&\hspace{2mm} +  \left( 
{\left[G_{ren}{}\!\,_{;}{}_{a}\right]}{}\!\,_{;}{}_{d} - 
    {\left[G_{ren}{}\!\,_{;}{}_{a}{}_{d}\right]} \right) \,
  \left( {\left[G_{ren}{}\!\,_{;}{}_{b}\right]}{}\!\,_{;}{}_{c} - 
    {\left[G_{ren}{}\!\,_{;}{}_{b}{}_{c}\right]} \right)    \cr && \left.
  \hspace{2mm} + \left( {\left[G_{ren}{}\!\,_{;}{}_{d}\right]}{}\!\,_{;}{}_{a} 
- 
    {\left[G_{ren}{}\!\,_{;}{}_{a}{}_{d}\right]} \right) \,
  \left( {\left[G_{ren}{}\!\,_{;}{}_{c}\right]}{}\!\,_{;}{}_{b} - 
    {\left[G_{ren}{}\!\,_{;}{}_{b}{}_{c}\right]} \right)    \right\} \cr
%
&& +  \left\{
   \left( -{\left[G_{ren}{}\!\,_{;}{}_{a}\right]}{}\!\,_{;}{}_{b} - 
    {\left[G_{ren}{}\!\,_{;}{}_{b}\right]}{}\!\,_{;}{}_{a} + 
    {\left[G_{ren}\right]}{}\!\,_{;}{}_{a}{}_{b} + 
    {\left[G_{ren}{}\!\,_{;}{}_{a}{}_{b}\right]} \right) \,
  {\left[G_{ren}{}\!\,_{;}{}_{c}{}_{d}\right]} \right.  \cr
&&\hspace{8mm} +  {\left[G_{ren}{}\!\,_{;}{}_{a}{}_{b}\right]}\,
  \left( -{\left[G_{ren}{}\!\,_{;}{}_{c}\right]}{}\!\,_{;}{}_{d} - 
    {\left[G_{ren}{}\!\,_{;}{}_{d}\right]}{}\!\,_{;}{}_{c} + 
    {\left[G_{ren}\right]}{}\!\,_{;}{}_{c}{}_{d} + 
    {\left[G_{ren}{}\!\,_{;}{}_{c}{}_{d}\right]} \right)    \cr
&&\hspace{8mm} +  {\left[G_{ren}\right]}\,\left( 
-{\left[G_{ren}{}\!\,_{;}{}_{a}{}_{b}{}_{c}
       \right]}{}\!\,_{;}{}_{d} - 
    {\left[G_{ren}{}\!\,_{;}{}_{a}{}_{b}{}_{d}\right]}{}\!\,_{;}{}_{c} + 
    {\left[G_{ren}{}\!\,_{;}{}_{a}{}_{b}\right]}{}\!\,_{;}{}_{c}{}_{d} + 
    {\left[G_{ren}{}\!\,_{;}{}_{a}{}_{b}{}_{d}{}_{c}\right]} \right)    \cr && 
\left.
  \hspace{8mm} + {\left[G_{ren}\right]}\,\left( 
-{\left[G_{ren}{}\!\,_{;}{}_{c}{}_{d}{}_{a}
       \right]}{}\!\,_{;}{}_{b} - 
    {\left[G_{ren}{}\!\,_{;}{}_{c}{}_{d}{}_{b}\right]}{}\!\,_{;}{}_{a} + 
    {\left[G_{ren}{}\!\,_{;}{}_{c}{}_{d}\right]}{}\!\,_{;}{}_{a}{}_{b} + 
    {\left[G_{ren}{}\!\,_{;}{}_{c}{}_{d}{}_{b}{}_{a}\right]} \right)    
\right\} \cr
%
&& -2 \left\{
   \left( {\left[G_{ren}\right]}{}\!\,_{;}{}_{a} - 
    {\left[G_{ren}{}\!\,_{;}{}_{a}\right]} \right) \,
  \left( {\left[G_{ren}{}\!\,_{;}{}_{c}{}_{d}\right]}{}\!\,_{;}{}_{b} - 
    {\left[G_{ren}{}\!\,_{;}{}_{c}{}_{d}{}_{b}\right]} \right)  \right.  \cr
&&\hspace{4mm} +  \left( {\left[G_{ren}\right]}{}\!\,_{;}{}_{b} - 
    {\left[G_{ren}{}\!\,_{;}{}_{b}\right]} \right) \,
  \left( {\left[G_{ren}{}\!\,_{;}{}_{c}{}_{d}\right]}{}\!\,_{;}{}_{a} - 
    {\left[G_{ren}{}\!\,_{;}{}_{c}{}_{d}{}_{a}\right]} \right)    \cr
&&\hspace{4mm} +  \left( {\left[G_{ren}\right]}{}\!\,_{;}{}_{c} - 
    {\left[G_{ren}{}\!\,_{;}{}_{c}\right]} \right) \,
  \left( {\left[G_{ren}{}\!\,_{;}{}_{a}{}_{b}\right]}{}\!\,_{;}{}_{d} - 
    {\left[G_{ren}{}\!\,_{;}{}_{a}{}_{b}{}_{d}\right]} \right)    \cr
&&\hspace{4mm} +  \left( {\left[G_{ren}\right]}{}\!\,_{;}{}_{d} - 
    {\left[G_{ren}{}\!\,_{;}{}_{d}\right]} \right) \,
  \left( {\left[G_{ren}{}\!\,_{;}{}_{a}{}_{b}\right]}{}\!\,_{;}{}_{c} - 
    {\left[G_{ren}{}\!\,_{;}{}_{a}{}_{b}{}_{c}\right]} \right)    \cr
&&\hspace{7mm} +  {\left[G_{ren}{}\!\,_{;}{}_{a}\right]}\,
  \left( -{\left[G_{ren}{}\!\,_{;}{}_{b}{}_{c}\right]}{}\!\,_{;}{}_{d} - 
    {\left[G_{ren}{}\!\,_{;}{}_{b}{}_{d}\right]}{}\!\,_{;}{}_{c} + 
    {\left[G_{ren}{}\!\,_{;}{}_{b}\right]}{}\!\,_{;}{}_{c}{}_{d} + 
    {\left[G_{ren}{}\!\,_{;}{}_{b}{}_{d}{}_{c}\right]} \right)    \cr
&&\hspace{7mm} +  {\left[G_{ren}{}\!\,_{;}{}_{b}\right]}\,
  \left( -{\left[G_{ren}{}\!\,_{;}{}_{a}{}_{c}\right]}{}\!\,_{;}{}_{d} - 
    {\left[G_{ren}{}\!\,_{;}{}_{a}{}_{d}\right]}{}\!\,_{;}{}_{c} + 
    {\left[G_{ren}{}\!\,_{;}{}_{a}\right]}{}\!\,_{;}{}_{c}{}_{d} + 
    {\left[G_{ren}{}\!\,_{;}{}_{a}{}_{d}{}_{c}\right]} \right)    \cr
&&\hspace{7mm} +  {\left[G_{ren}{}\!\,_{;}{}_{c}\right]}\,
  \left( -{\left[G_{ren}{}\!\,_{;}{}_{a}{}_{d}\right]}{}\!\,_{;}{}_{b} - 
    {\left[G_{ren}{}\!\,_{;}{}_{b}{}_{d}\right]}{}\!\,_{;}{}_{a} + 
    {\left[G_{ren}{}\!\,_{;}{}_{d}\right]}{}\!\,_{;}{}_{a}{}_{b} + 
    {\left[G_{ren}{}\!\,_{;}{}_{b}{}_{d}{}_{a}\right]} \right)    \cr && \left.
  \hspace{7mm} + {\left[G_{ren}{}\!\,_{;}{}_{d}\right]}\,
  \left( -{\left[G_{ren}{}\!\,_{;}{}_{a}{}_{c}\right]}{}\!\,_{;}{}_{b} - 
    {\left[G_{ren}{}\!\,_{;}{}_{b}{}_{c}\right]}{}\!\,_{;}{}_{a} + 
    {\left[G_{ren}{}\!\,_{;}{}_{c}\right]}{}\!\,_{;}{}_{a}{}_{b} + 
    {\left[G_{ren}{}\!\,_{;}{}_{b}{}_{c}{}_{a}\right]} \right)    \right\} \cr
%
&& + 2 \left\{
 \left( \left( {\left[G_{ren}\right]}{}\!\,_{;}{}_{a} - 
       {\left[G_{ren}{}\!\,_{;}{}_{a}\right]} \right) \,
     \left( {\left[G_{ren}\right]}{}\!\,_{;}{}_{b} - 
       {\left[G_{ren}{}\!\,_{;}{}_{b}\right]} \right)  + 
    {\left[G_{ren}{}\!\,_{;}{}_{a}\right]}\,
     {\left[G_{ren}{}\!\,_{;}{}_{b}\right]} \right) \,{R{}_{c}{}_{d}}
\right. \cr
%
&& \hspace{4mm} + \left. 
 \left( \left( {\left[G_{ren}\right]}{}\!\,_{;}{}_{c} - 
       {\left[G_{ren}{}\!\,_{;}{}_{c}\right]} \right) \,
     \left( {\left[G_{ren}\right]}{}\!\,_{;}{}_{d} - 
       {\left[G_{ren}{}\!\,_{;}{}_{d}\right]} \right)  + 
    {\left[G_{ren}{}\!\,_{;}{}_{c}\right]}\,
     {\left[G_{ren}{}\!\,_{;}{}_{d}\right]} \right) \,{R{}_{a}{}_{b}}
\right\} \cr
%
&& + 
\left\{
 \left( {\left[G_{ren}{}\!\,_{;}{}_{a}\right]}{}\!\,_{;}{}_{b} + 
    {\left[G_{ren}{}\!\,_{;}{}_{b}\right]}{}\!\,_{;}{}_{a} - 
    {\left[G_{ren}\right]}{}\!\,_{;}{}_{a}{}_{b} - 
    2\,{\left[G_{ren}{}\!\,_{;}{}_{a}{}_{b}\right]} \right) \,{R{}_{c}{}_{d}}
\right. \cr
%
&& \hspace{2mm} + \left. 
 \left( {\left[G_{ren}{}\!\,_{;}{}_{c}\right]}{}\!\,_{;}{}_{d} + 
    {\left[G_{ren}{}\!\,_{;}{}_{d}\right]}{}\!\,_{;}{}_{c} - 
    {\left[G_{ren}\right]}{}\!\,_{;}{}_{c}{}_{d} - 
    2\,{\left[G_{ren}{}\!\,_{;}{}_{c}{}_{d}\right]} \right) \,{R{}_{a}{}_{b}}
\right. \cr
%
&& \hspace{2mm} + \left. 
 {R{}_{a}{}_{b}}\,{R{}_{c}{}_{d}} {\left[G_{ren}\right]} 
\right\}  {\left[G_{ren}\right]}
\end{eqnarray}
\begin{eqnarray}
 144 N'_{ab} &=& 
%
-4 \left(
 \left( {\left[G_{ren}{}\!\,_{;}{}_{b}\right]}{}\!\,_{;}{}_{p} - 
    {\left[G_{ren}{}\!\,_{;}{}_{p}{}_{b}\right]} \right) \,
  \left( {\left[G_{ren}{}\!\,_{;}{}_{a}\right]}{}\!\,^{;}{}^{p} - 
    {\left[G_{ren}{}\!\,_{;}{}_{a}{}^{p}\right]} \right)  
\right. \cr && \left. \hspace{3mm} +
 \left( {\left[G_{ren}{}\!\,_{;}{}_{p}\right]}{}\!\,_{;}{}_{b} - 
    {\left[G_{ren}{}\!\,_{;}{}_{p}{}_{b}\right]} \right) \,
  \left( {\left[G_{ren}{}\!\,^{;}{}^{p}\right]}{}\!\,_{;}{}_{a} - 
    {\left[G_{ren}{}\!\,_{;}{}_{a}{}^{p}\right]} \right)  
\right) \cr
%
&&+4 \left(
 \left( {\left[G_{ren}\right]}{}\!\,_{;}{}_{a} - 
    {\left[G_{ren}{}\!\,_{;}{}_{a}\right]} \right) \,
  \left( {\left[G_{ren}{}\!\,_{;}{}_{p}{}^{p}\right]}{}\!\,_{;}{}_{b} - 
    {\left[G_{ren}{}\!\,_{;}{}_{p}{}^{p}{}_{b}\right]} \right)  
\right. \cr && \left. \hspace{3mm} +
 \left( {\left[G_{ren}\right]}{}\!\,_{;}{}_{b} - 
    {\left[G_{ren}{}\!\,_{;}{}_{b}\right]} \right) \,
  \left( {\left[G_{ren}{}\!\,_{;}{}_{p}{}^{p}\right]}{}\!\,_{;}{}_{a} - 
    {\left[G_{ren}{}\!\,_{;}{}_{p}{}^{p}{}_{a}\right]} \right)  
\right. \cr && \left. \hspace{5mm} +
 {\left[G_{ren}{}\!\,_{;}{}_{a}\right]}\,
  \left( -{\left[G_{ren}{}\!\,_{;}{}_{p}{}_{b}\right]}{}\!\,^{;}{}^{p} - 
    {\left[G_{ren}{}\!\,_{;}{}_{b}{}^{p}\right]}{}\!\,_{;}{}_{p} + 
    {\left[G_{ren}{}\!\,_{;}{}_{b}\right]}{}\!\,_{;}{}_{p}{}^{p} + 
    {\left[G_{ren}{}\!\,_{;}{}_{b}{}^{p}{}_{p}\right]} \right)  
\right. \cr && \left. \hspace{5mm} +
 {\left[G_{ren}{}\!\,_{;}{}_{b}\right]}\,
  \left( -{\left[G_{ren}{}\!\,_{;}{}_{p}{}_{a}\right]}{}\!\,^{;}{}^{p} - 
    {\left[G_{ren}{}\!\,_{;}{}_{a}{}^{p}\right]}{}\!\,_{;}{}_{p} + 
    {\left[G_{ren}{}\!\,_{;}{}_{a}\right]}{}\!\,_{;}{}_{p}{}^{p} + 
    {\left[G_{ren}{}\!\,_{;}{}_{a}{}^{p}{}_{p}\right]} \right)  
\right) \cr
&&2 \left(
 {\left[G_{ren}{}\!\,^{;}{}^{p}\right]}\,
  \left( -{\left[G_{ren}{}\!\,_{;}{}_{p}{}_{a}\right]}{}\!\,_{;}{}_{b} - 
    {\left[G_{ren}{}\!\,_{;}{}_{p}{}_{b}\right]}{}\!\,_{;}{}_{a} + 
    {\left[G_{ren}{}\!\,_{;}{}_{p}\right]}{}\!\,_{;}{}_{a}{}_{b} + 
    {\left[G_{ren}{}\!\,_{;}{}_{p}{}_{b}{}_{a}\right]} \right)  
\right. \cr && \left. \hspace{1mm} +
 \left( {\left[G_{ren}\right]}{}\!\,^{;}{}^{p} - 
    {\left[G_{ren}{}\!\,^{;}{}^{p}\right]} \right) \,
  \left( {\left[G_{ren}{}\!\,_{;}{}_{a}{}_{b}\right]}{}\!\,_{;}{}_{p} - 
    {\left[G_{ren}{}\!\,_{;}{}_{a}{}_{b}{}_{p}\right]} \right)  
\right) \cr
%
&& -\left( R\,{\left[G_{ren}\right]}\,
    \left( {\left[G_{ren}{}\!\,_{;}{}_{a}\right]}{}\!\,_{;}{}_{b} + 
      {\left[G_{ren}{}\!\,_{;}{}_{b}\right]}{}\!\,_{;}{}_{a} - 
      {\left[G_{ren}\right]}{}\!\,_{;}{}_{a}{}_{b} - 
      2\,{\left[G_{ren}{}\!\,_{;}{}_{a}{}_{b}\right]} \right)  \right)  \cr
%
&&  -\left( \left( {\left[G_{ren}\right]}{}\!\,_{;}{}_{p}\,
       {\left[G_{ren}\right]}{}\!\,^{;}{}^{p} - 
      2\,{\left[G_{ren}\right]}{}\!\,_{;}{}_{p}\,
       {\left[G_{ren}{}\!\,^{;}{}^{p}\right]} + 
      2\,{\left[G_{ren}{}\!\,_{;}{}_{p}\right]}\,
       {\left[G_{ren}{}\!\,^{;}{}^{p}\right]} \right) \,{R{}_{a}{}_{b}}
    \right)  \cr
%
&& -2\,{\left[G_{ren}\right]}\,\left( 2\,
     {\left[G_{ren}{}\!\,_{;}{}_{p}\right]}{}\!\,^{;}{}^{p} - 
    {\left[G_{ren}\right]}{}\!\,_{;}{}_{p}{}^{p} - 
    2\,{\left[G_{ren}{}\!\,_{;}{}_{p}{}^{p}\right]} \right) \,{R{}_{a}{}_{b}} 
\cr
%
&& -\left( R\,{R{}_{a}{}_{b}} \right)  {{{\left[G_{ren}\right]}}^2}
\end{eqnarray}
\begin{eqnarray}
 288 N'''&=& 
%
2 \left(
 \left( {\left[G_{ren}{}\!\,_{;}{}_{p}\right]}{}\!\,_{;}{}_{q} - 
    {\left[G_{ren}{}\!\,_{;}{}_{p}{}_{q}\right]} \right) \,
  \left( {\left[G_{ren}{}\!\,^{;}{}^{p}\right]}{}\!\,^{;}{}^{q} - 
    {\left[G_{ren}{}\!\,^{;}{}^{p}{}^{q}\right]} \right) 
\right. \cr && \left. +
 \left( {\left[G_{ren}{}\!\,_{;}{}_{q}\right]}{}\!\,_{;}{}_{p} - 
    {\left[G_{ren}{}\!\,_{;}{}_{p}{}_{q}\right]} \right) \,
  \left( {\left[G_{ren}{}\!\,^{;}{}^{q}\right]}{}\!\,^{;}{}^{p} - 
    {\left[G_{ren}{}\!\,^{;}{}^{p}{}^{q}\right]} \right) 
\right) \cr
%
&&-8 \left(
 \left( 2\,{\left[G_{ren}{}\!\,_{;}{}_{p}\right]}{}\!\,^{;}{}^{p} - 
    {\left[G_{ren}\right]}{}\!\,_{;}{}_{p}{}^{p} - 
    {\left[G_{ren}{}\!\,_{;}{}_{p}{}^{p}\right]} \right) \,
  {\left[G_{ren}{}\!\,_{;}{}_{q}{}^{q}\right]}
\right. \cr && \left. \hspace{12mm} 
 {\left[G_{ren}\right]}\,\left( 2\,
     {\left[G_{ren}{}\!\,_{;}{}_{p}{}^{p}{}_{q}\right]}{}\!\,^{;}{}^{q} - 
    {\left[G_{ren}{}\!\,_{;}{}_{p}{}^{p}\right]}{}\!\,_{;}{}_{q}{}^{q} - 
    {\left[G_{ren}{}\!\,_{;}{}_{p}{}^{p}{}_{q}{}^{q}\right]} \right) 
\right) \cr
%
&& + 8 \left(
\left( {\left[G_{ren}\right]}{}\!\,_{;}{}_{p} - 
    {\left[G_{ren}{}\!\,_{;}{}_{p}\right]} \right) \,
  {\left[G_{ren}{}\!\,_{;}{}_{q}{}^{q}{}^{p}\right]}
- {\left[G_{ren}\right]}{}\!\,_{;}{}_{p}\,
  {\left[G_{ren}{}\!\,_{;}{}_{q}{}^{q}\right]}{}\!\,^{;}{}^{p}
\right. \cr && \left. \hspace{3mm} + 
 {\left[G_{ren}{}\!\,_{;}{}_{p}\right]}\,
  \left( 2\,{\left[G_{ren}{}\!\,_{;}{}_{q}{}^{p}\right]}{}\!\,^{;}{}^{q} + 
    {\left[G_{ren}{}\!\,_{;}{}_{q}{}^{q}\right]}{}\!\,^{;}{}^{p} - 
    {\left[G_{ren}{}\!\,^{;}{}^{p}\right]}{}\!\,_{;}{}_{q}{}^{q} - 
    {\left[G_{ren}{}\!\,_{;}{}_{q}{}^{p}{}^{q}\right]} \right) 
\right) \cr
%
&&    2\,R\,\left( {\left[G_{ren}\right]}{}\!\,_{;}{}_{p}\,
     {\left[G_{ren}\right]}{}\!\,^{;}{}^{p} - 
    2\,{\left[G_{ren}\right]}{}\!\,_{;}{}_{p}\,
     {\left[G_{ren}{}\!\,^{;}{}^{p}\right]} + 
    2\,{\left[G_{ren}{}\!\,_{;}{}_{p}\right]}\,
     {\left[G_{ren}{}\!\,^{;}{}^{p}\right]} \right)  \cr
%
&& +  4\,R\,\left( 2\,{\left[G_{ren}{}\!\,_{;}{}_{p}\right]}{}\!\,^{;}{}^{p} - 
    {\left[G_{ren}\right]}{}\!\,_{;}{}_{p}{}^{p} - 
    2\,{\left[G_{ren}{}\!\,_{;}{}_{p}{}^{p}\right]} \right)  
{\left[G_{ren}\right]} \cr
%
&& + {R^2} {{{\left[G_{ren}\right]}}^2}
\end{eqnarray}
\end{mathletters}

%
%
For the minimal coupling ($\xi=0$) case:
\begin{mathletters}
\begin{eqnarray}
8 N'_{abcd}&=& 
%
\hspace{4mm}   \left( {\left[G_{ren}{}\!\,_{;}{}_{a}\right]}{}\!\,_{;}{}_{c} - 
    {\left[G_{ren}{}\!\,_{;}{}_{a}{}_{c}\right]} \right) \,
  \left( {\left[G_{ren}{}\!\,_{;}{}_{b}\right]}{}\!\,_{;}{}_{d} - 
    {\left[G_{ren}{}\!\,_{;}{}_{b}{}_{d}\right]} \right)    \cr
&&   + \left( {\left[G_{ren}{}\!\,_{;}{}_{c}\right]}{}\!\,_{;}{}_{a} - 
    {\left[G_{ren}{}\!\,_{;}{}_{a}{}_{c}\right]} \right) \,
  \left( {\left[G_{ren}{}\!\,_{;}{}_{d}\right]}{}\!\,_{;}{}_{b} - 
    {\left[G_{ren}{}\!\,_{;}{}_{b}{}_{d}\right]} \right)   \cr
&&   + \left( {\left[G_{ren}{}\!\,_{;}{}_{a}\right]}{}\!\,_{;}{}_{d} - 
    {\left[G_{ren}{}\!\,_{;}{}_{a}{}_{d}\right]} \right) \,
  \left( {\left[G_{ren}{}\!\,_{;}{}_{b}\right]}{}\!\,_{;}{}_{c} - 
    {\left[G_{ren}{}\!\,_{;}{}_{b}{}_{c}\right]} \right)   \cr
&&     + \left( {\left[G_{ren}{}\!\,_{;}{}_{d}\right]}{}\!\,_{;}{}_{a} - 
    {\left[G_{ren}{}\!\,_{;}{}_{a}{}_{d}\right]} \right) \,
  \left( {\left[G_{ren}{}\!\,_{;}{}_{c}\right]}{}\!\,_{;}{}_{b} - 
    {\left[G_{ren}{}\!\,_{;}{}_{b}{}_{c}\right]} \right) 
\end{eqnarray}
\begin{eqnarray}
8 N'_{ab}&=& 
%
- \left( {\left[G_{ren}{}\!\,_{;}{}_{b}\right]}{}\!\,_{;}{}_{p} - 
    {\left[G_{ren}{}\!\,_{;}{}_{p}{}_{b}\right]} \right) \,
  \left( {\left[G_{ren}{}\!\,_{;}{}_{a}\right]}{}\!\,^{;}{}^{p} - 
    {\left[G_{ren}{}\!\,_{;}{}_{a}{}^{p}\right]} \right)  \cr
&& - \left( {\left[G_{ren}{}\!\,_{;}{}_{p}\right]}{}\!\,_{;}{}_{b} - 
    {\left[G_{ren}{}\!\,_{;}{}_{p}{}_{b}\right]} \right) \,
  \left( {\left[G_{ren}{}\!\,^{;}{}^{p}\right]}{}\!\,_{;}{}_{a} - 
    {\left[G_{ren}{}\!\,_{;}{}_{a}{}^{p}\right]} \right)  \cr
&& - {m^2}\,\left( {\left[G_{ren}\right]}{}\!\,_{;}{}_{a} - 
    {\left[G_{ren}{}\!\,_{;}{}_{a}\right]} \right) \,
  \left( {\left[G_{ren}\right]}{}\!\,_{;}{}_{b} - 
    {\left[G_{ren}{}\!\,_{;}{}_{b}\right]} \right)  \cr
&& - {m^2}\,{\left[G_{ren}{}\!\,_{;}{}_{a}\right]}\,
  {\left[G_{ren}{}\!\,_{;}{}_{b}\right]}
\end{eqnarray}
\begin{eqnarray}
8 N'''&=& 
%
{\frac{1}{2}} \left(
 \left( {\left[G_{ren}{}\!\,_{;}{}_{p}\right]}{}\!\,_{;}{}_{q} - 
    {\left[G_{ren}{}\!\,_{;}{}_{p}{}_{q}\right]} \right) \,
  \left( {\left[G_{ren}{}\!\,^{;}{}^{p}\right]}{}\!\,^{;}{}^{q} - 
    {\left[G_{ren}{}\!\,^{;}{}^{p}{}^{q}\right]} \right) 
\right. \cr && \left. \hspace{2mm}+
 \left( {\left[G_{ren}{}\!\,_{;}{}_{q}\right]}{}\!\,_{;}{}_{p} - 
    {\left[G_{ren}{}\!\,_{;}{}_{p}{}_{q}\right]} \right) \,
  \left( {\left[G_{ren}{}\!\,^{;}{}^{q}\right]}{}\!\,^{;}{}^{p} - 
    {\left[G_{ren}{}\!\,^{;}{}^{p}{}^{q}\right]} \right) 
\right) \cr
%
&& +   {m^2}\,\left( {\left[G_{ren}\right]}{}\!\,_{;}{}_{p}\,
     {\left[G_{ren}\right]}{}\!\,^{;}{}^{p} - 
    2\,{\left[G_{ren}\right]}{}\!\,_{;}{}_{p}\,
     {\left[G_{ren}{}\!\,^{;}{}^{p}\right]} + 
    2\,{\left[G_{ren}{}\!\,_{;}{}_{p}\right]}\,
     {\left[G_{ren}{}\!\,^{;}{}^{p}\right]} \right)  \cr
%
&& + {m^4} {{{\left[G_{ren}\right]}}^2}
\end{eqnarray}
\end{mathletters}

\section{Further Developments}

In this article we have derived a general expression for
the noise kernel, or the vacuum expectation value of the
stress energy bi-tensor for a quantum scalar field in 
a general curved space time  using the point separation 
method. The general form is expressed as products of covariant derivatives of 
the
quantum field's Green function. It  is finite when
the noise kernel is evaluated for distinct pairs of points
(and non-null points for a massless field). We also
have shown the trace of the noise kernel  vanishes, confirming there is no noise
associated with the trace anomaly. This holds regardless
of issues of regularization of the noise kernel.

The noise kernel as a two point function of the
stress energy tensor diverges  as the
pair of points are brought together, representing 
the ``standard'' ultraviolet divergence present in the
quantum field theory. By using the modified point
separation regularization method we render the
field's Green function finite in the coincident limit.
This in turn permits the derivation of the formal expression 
for the regularized coincident limit of the noise kernel.

When the Green function is available in closed analytic
form one can carry out an  end point expansion according
to (\ref{general-endpt-series}),  displaying  the ultraviolet
divergence. Subtraction of the  Hadamard 
ansatz (\ref{ref-HadamardAnsatz}), expressed as a series 
expansion, (\ref{ref-vwseries}), will render this Green function
finite in the coincident limit. With this, one can calculate the noise
kernel for a variety of spacetimes, as will be developed in Papers II, III.
Here we give an outline of the program in this series.

A common analytic approximation is the Gaussian \cite{BekPark81},
which, in ultrastatic spacetimes,  provides a closed form expression for 
the Green function. In Paper II we make
use of this form to evaluate the noise kernel for a massless scalar
field in hot flat space and the Einstein universe. For
hot flat space, the Gaussian Green function is exact, while for
the Einstein universe, the exact Green function is known.
We also work explicitly with the ultrastatic metric conformal to the
Schwarzschild metric. Though the approximate Green function is
known to be a fairly good approximation for the stress tensor, {\it i.e.},
to second order, we find the approximation breaks down at fourth
order (the noise kernel needs up to four covariant derivatives of the
Green function). This manifest in  the noise kernel trace 
computed under Gaussian approximation failing to vanish;
indeed the magnitude of noise kernel  becomes comparable
to the components themselves. This form of the  Green function fails
to satisfy the field equations at fourth order.

In Paper III we take advantage of the simple conformal transformation property
of the scalar field's Green function to compute the
noise kernel for a thermal fields in a flat FRW Universe as it is obtainable 
from that of hot flat space by a time dependent scale factor, and likewise
use the Einstein Universe result to obtain the noise kernel  for a closed
FRW Universe. The most interesting case is when we conformally
transform to the Schwarzschild metric. Though our results based on
the Gaussian approximation break down  close to
the event horizon, we can obtain reasonable results for the fluctuations of the
stress tensor of the Hawking flux in the far field region and check with 
analytic
results \cite{CamHu}. Along the
way, we verify our procedure by explicitly re-deriving the
Page result \cite{Page82} for the stress tensor. We note that in Page's
original work, the direct use of the conformal transformation was
circumvented by ``guessing'' the solution to a functional
differential equation. Our result is the first we know where
the methodology of point separation was carried all the way
through to the final result. That we get the known results
is a check on our method and its correct implementation.

We presently outline how the work presented in this article are used to
compute the coincident limit of the noise kernel. In future
articles in this series we will present the details and results.
We assume we have an analytic closed form expression for the
Green function; we focus on the Gaussian 
approximation \cite{BekPark81}. As mentioned above, this
Green function is expanded in an end point series and the
Hadamard ansatz is subtracted, generating the 
series expansion of the renormalized Green function. At this
point we could directly substitute this expansion in our
expression for the coincident limit of the noise kernel
(\ref{coincident-noise}). The resulting expression becomes
quite large and it is hard to glean any physical meaning 
from the resulting expressions.
Instead we let this be the point in the problem when
an explicit metric is chosen. Once this is done, in the
context of the symbolic computation, it is straightforward
to determine the component values of the Green function
expansion tensors. For $d=4$, there is at most 70 unique
values. From this we can readily generate all the
needed component values of the coincident limits of the
covariant derivatives of the Green function, along with
the covariant derivatives of the coincident limits. It is
these explicitly evaluated tensors that are then substituted
in  (\ref{coincident-noise}) to get the final result.

We must stress that though all this is done on a computer,
no numerical approximations are used; all work is done
symbolically in terms of the explicit functional form
of the metric and the parameters of the field. The final
results are exact to the extent that the analytic form of
the Green function is exact.

As will be seen in Paper II and III the basic procedures
for generating the needed series expansions are recursive
on the expansion order. For the noise kernel we need results
up to fourth order in the separation distance. The well
established work for the stress tensor is to second order. 
This provides a check of our code by verifying we always
get the known results for the stress tensor expectation
value. Once we know the second order recursion is correct,
we know the algorithm is functioning as desired and the (new)
fourth order terms are correct. This becomes particularly
important when we consider metrics conformally related, as
we get intermediate results of up to 1100 terms in length.

When we are interested in spacetimes for which we know the
Green function in a conformally related space time, we
have modified the above procedure. The simple conformal
transformation property of Green functions allows us to get
the Green function in the physical metric we are interested in
from that  in the conformally related metric (for the
Gaussian approximation, this is the conformally optical metric).
The main obstacle to overcome is the subtraction of the 
Hadamard ansatz. The divergent Green function is defined in terms
of the optical metric while the Hadamard ansatz in terms of
the physical metric. We need to re-express the transformed
optical metric in terms of the physical metric. The defining
equations for the geometric objects
({\it e.g.} Eqns \ref{define-sigma} and \ref{define-VanD})
on the optical metric are transformed to the physical metric
and recursively solved. Now the Green function series
expansion can be written solely in terms of the physical
metric. It is at this point that  the symbolic computation 
environment reigns; the fourth order
term in the expansion of the renormalized Green function
has over 1100 terms. On the other hand, it is exactly
in this context that we re-derive the Page \cite{Page82}
result for the stress tensor vacuum expectation value
on the Scharzschild black hole. As we stressed above, since
all the series expansions used are recursively derived,
so with the lower order results checked,  we
know the resulting expression is correct. Now having the
general expansion of the renormalized Green function, we
proceed as we outlined above and choose a metric and
compute the coincident limit of the noise kernel.

Though our main focus in this article has been to lay
out the groundwork for computing the coincident limit of the
noise kernel, our result (\ref{general-noise-kernel}) is
completely general. This expression for the noise kernel
only assumes a scalar field and can be used with or without
first considering issues of renormalization of the Green
function. Also, our result for the coincident limit,
(\ref{coincident-noise}), holds regardless of the choice
of Green function or metric. It only requires that the
Green function had a meaningful coincident limit. 
It can become the starting point for numerical work
on the noise kernel, an avenue worthy of some deliberation. \\

%
\noindent {\bf Acknowledgement}  NGP thanks Professor R. M. Wald for 
correspondence 
on his  regularization procedures, and Professors S. Christensen and L. Parker 
for the use of
the MathTensor program. Preliminary result of this work was reported by BLH at 
the Third 
Peyresq meeting on quantum cosmology and stochastic gravity,  during which he 
enjoyed
the hospitality of Professor E. Gundzig and discussions with Professor E. 
Verdaguer.
This work is supported in part by NSF grant PHY98-00967.

\end{document}